\documentstyle[12pt,aasms4,psfig]{article} 
\begin{document} 
\def \grsim{\:^{>}_{\sim}\:}
\def \lesssim{\:^{<}_{\sim}\:}
\title{THE REDSHIFT-SPACE TWO-POINT CORRELATION FUNCTIONS OF GALAXIES AND 
GROUPS IN THE NEARBY OPTICAL GALAXY SAMPLE}
\author{Giuliano GIURICIN$^{1, 2}$, Srdjan SAMUROVI\'C$^{1, 3}$, Marisa 
GIRARDI$^{1}$,\\
Marino MEZZETTI$^{1}$, Christian MARINONI$^{1}$}
\affil{
$^{1}$ Dipartimento di Astronomia, Univ. di Trieste, via G. B. Tiepolo 11, 
34131 Trieste, Italy\\ 
$^{2}$ SISSA, via Beirut 4, 34013 Trieste, Italy\\
$^{3}$ International Center for Theoretical Physics, Strada Costiera 11, 34014 Trieste, Italy\\
E-mail: giuricin@sissa.it, srdjan@ts.astro.it, girardi@ts.astro.it\\
mezzetti@ts.astro.it, marinoni@sissa.it\\
}
\begin{abstract}
We use the two-point correlation function in redshift space, $\xi(s)$, to study the 
clustering of the galaxies and groups of the Nearby Optical Galaxy (NOG) sample, 
which is a nearly all-sky ($|b|>20^{\circ}$), complete, magnitude-limited sample of 
$\sim$7000 bright and nearby optical galaxies ($cz\leq 6000$ km/s).

The correlation function of galaxies is well described by a power law, $\xi(s)=
(s/s_0)^{-\gamma}$, with slope $\gamma\sim1.5$ and $s_0\sim6.4\;h^{-1}$Mpc (on scales 
$2.7 - 12 h^{-1}$Mpc), in substantial agreement with previous results of several   
redshift surveys of optical galaxies. 

Splitting NOG into different morphological subsamples, we confirm the existence of
morphological segregation between early- and late-type galaxies (out to
$20\;h^{-1}$Mpc) and, in particular, we find a gradual decreasing of the strength of
clustering from the S0 galaxies to the late-type spirals, on intermediate scales 
(around $5\;h^{-1}$Mpc). The relative bias factor between early- and late-type 
galaxies appears to be substantially constant with scale. 

Furthermore, luminous galaxies turn out to be more 
clustered than dim galaxies. The luminosity segregation, which  is significant for 
both early- and late-type objects, starts to become appreciable only 
for galaxies brighter than $M_B\sim -19.5 + 5 \log h$ ($\sim 0.6 L^*$) and is
independent on scale.

The NOG groups identified with the hierarchical and percolation algorithms show 
similar clustering properties, with a degree of clustering which is intermediate 
between galaxies and clusters. The group correlation functions are characterized 
by  $s_0$-values ranging from $\sim 8\;h^{-1}$ Mpc (for groups with at least 
three members) to $\sim10\;h^{-1}$ Mpc (for groups with at least
five members). The degree of group clustering depends on the physical properties of
groups. Specifically, groups with greater velocity dispersions, sizes and masses 
tend to be more clustered than those with lower values of these quantities. 

\end{abstract}


\section{Introduction} 

The two-point correlation function (hereafter CF) has been the first and most widely used approach
to quantify the degree of clustering in a galaxy sample (Totsuji \& Kihara 1969;
Peebles 1980). This statistics has been widely applied to a variety of samples of
optically- and infrared-selected galaxies, groups and clusters of galaxies. Besides,
it has been used to characterize the dependence of the galaxy clustering on the
internal properties of galaxies such as morphology (e.g., Davis \& Geller 1976),
color (e.g., Tucker et al. 1995), surface brightness (Davis \& Djorgovski 1985),
luminosity (e.g., Maurogordato \& Lachi\`eze-Rey 1987), and internal dynamics 
(e.g., White, Tully, \& Davis 1988). 

From galaxies to clusters the CF in redshift space,
$\xi(s)$, is characterized by a power-law form, $\xi(s) = (s/s_0)^{-\gamma}$, with
$\gamma\sim 1.5 - 2$ for a variety of systems. The correlation length $s_0$ ranges
from $\sim 5 - 7.5\;h^{-1}$ Mpc for optically-selected galaxies (e. g., Willmer, 
da Costa, \& Pellegrini 1998 and references cited therein) 
to $s_0\grsim 15\;h^{-1}$ Mpc for optically-
and X-ray selected galaxy clusters (e.g., Bahcall \& Soneira 1983; Postman, Huchra, 
\& Geller 1991; Peacock \& West 1992; Croft et al. 1997; Borgani, Plionis, \& Kolokotronis
1999). (Throughout, the Hubble constant is $H_0=100\;h$ km~s$^{-1}$ Mpc$^{-1}$). 

Results are far less clear for loose galaxy groups. Jing \& Zhang (1988) and Maia \&
da Costa (1989) calculated the CF of the groups identified in the
CfA1 and SSRS redshift surveys and found that the amplitude of the group CF 
is lower than that of the galaxy CF. Later, from the
analysis of the groups identified in a subsample of the CfA2 redshift survey,
Ramella, Geller, \& Huchra (1990) found that the amplitudes of the two CFs 
were similar (see also Kalinkov \& Kuneva 1990). Recently, Trasarti-Battistoni, Invernizzi,  
\& Bonometto (1997), who analyzed groups identified in the Perseus-Pisces 
region, and Girardi, Boschin, \& da Costa (2000), who considered the 
groups respectively identified by Ramella, Pisani, \& Geller (1997) and Ramella et 
al. (2000) in the CfA2 and SSRS2 redshift surveys, stressed that groups are 
more clustered than galaxies. 

In this paper we determine the CF  (in redshift space) of
the galaxies and groups in the nearby universe, by using the NOG ("Nearby Optical
Galaxies") sample (Giuricin et al. 2000), which is a complete, distance-limited 
($cz\leq 6000$ km/s) and magnitude-limited sample of $\sim$7000 nearby and bright 
optical galaxies, which covers $\sim2/3$ of the sky (8.27 sr). 

This paper is the fourth in a series of papers, in which we investigate on the
properties of the large-scale galaxy distribution in the nearby universe by using a 
nearly all-sky sample of optical galaxies (Marinoni et al. 1998 , Paper I; Marinoni et al.
1999, Paper II; Giuricin et al. 2000, Paper III). In a forthcoming paper 
(Marinoni, Hudson, \& Giuricin 2001) the optical luminosity function of all galactic systems   
derived from the NOG will be compared to theoretical predictions of the  mass 
function of virialized haloes. 

The NOG is currently one of the largest samples used in the determination of the 
galaxy CF, and, compared to other wide-angle, comparatively 
shallow, bright magnitude-limited galaxy samples, such as the CfA1, CfA2, 
and SSRS2 redshift surveys, the NOG may be less sensitive to local density 
fluctuations because it covers a much larger solid angle.
  
Besides comparing our results with those obtained from other galaxy samples, with
our wide sample we examine the dependence of the correlation properties of
galaxies on galaxy morphology, attempting a subdivision of galaxies in several 
morphological subtypes.  

We also investigate on the dependence of the galaxy CF  
on galaxy luminosity. In this respect our analysis 
differs from previous relevant studies because we rely on blue total 
magnitudes fully corrected for internal extinction, Galactic extinction, and K-dimming. 
These corrections lead to brighter magnitudes. In particular the first correction, 
which is conspicuous in very inclined spirals, is generally neglected in generic 
redshift surveys, which comprise many faint galaxies of unknown inclination.
Some authors (e.g. Hasegawa \& Umemura 1993) have contended that this 
correction has a non negligible impact on the determination of luminosity 
segregation.   

Furthermore, we investigate on the
dependence of the group CF on group properties, relying on two
catalogs of groups identified with the hierarchical and percolation {\it friends of
friends} algorithms.  Being extracted from the same galaxy sample, these catalogs 
allow us to investigate, for the first time, on possible differences in clustering 
properties strictly related to differences in the algorithm adopted.  
In the literature there are as yet no studies dealing
with the CF of groups identified with the hierarchical algorithm.

The outline of our paper is as follows. In \S 2 we briefly describe the NOG samples 
of galaxies and groups used in this study. In \S 3 we deal with  the estimate of the 
redshift-space CF. In \S 4 we present the CF of the
NOG galaxies. In \S 5 and \S6 we examine the dependence of the galaxy CF  
on galaxy morphology and luminosity, respectively. In \S 7 we present the 
CF of the NOG groups and compare it with that of galaxies in order 
to determine the relative clustering properties. In \S 8 we explore the dependence 
of the group CF on group properties (e.g., on velocity dispersions, 
radii, and masses). Conclusions are drawn in \S 9.

\section{The Samples of Galaxies and Groups}

In this work we use the NOG sample of
galaxies (described in Paper III) basically extracted from the
Lyon-Meudon Extragalactic Database (LEDA; e.g. Paturel et al. 1997) according to the
following selection criteria: i) galactic latitudes $|b|>20^{\circ}$; ii) recession
velocities (evaluated in the Local Group rest frame) $cz\leq 6000$ km/s; iii) corrected
total blue magnitudes $B\leq 14$ mag. In the LEDA compilation, which collected and
homogenized several data for all the galaxies of the main optical catalogs, the
original raw data (blue apparent magnitudes and angular sizes) have been transformed
to the standard systems of the RC3 catalog (de Vaucouleurs et al. 1991) and have
been corrected for Galactic extinction, internal extinction, and K-dimming. The
degree of the redshift completeness of the NOG is estimated to be 97\%. In this work
we consider the sample of 7028 galaxies having $cz\geq 50$ km/s.  

Almost all NOG galaxies (98.7\%) have a morphological classification.
We divide NOG galaxies into two broad morphological bins, early-type galaxies, i.e.  E-S0 galaxies  
($T<-1.5$) and late-type galaxies ($T\geq-1.5$), hereafter denoted as Sp, which comprise
S0/a galaxies and later types  
(T is the morphological type code system of the RC3 catalog). 
We also attempt a finer morphological subdivision, dividing galaxies into 6 morphological bins,
E ($T<-2.5$), S0 ($-2.5\leq T<-1.5$), S0/a ($-1.5\leq T<0.5$), Sa 
($0.5\leq T<2.5$), Sb ($2.5\leq T<3.5$), and later types (Sc$+$Sd$+$Sm$+$Irr, i.e.$T\geq3.5$,  
hereafter simply denoted as Scd). The first subsample, hereafter denoted by E for the sake of 
simplicity, does not comprise only ellipticals but also lenticulars, since it contains also
objects broadly classified as E-S0.

We also use the catalogs of NOG groups identified within the NOG by means of the 
hierarchical (H) and the percolation (P) algorithms (see Paper III). 

In the H method the authors adopted the galaxy luminosity density as the affinity
parameter used to construct the dendrogram and cut the hierarchy at a luminosity
density corresponding to a luminosity
density contrast of 45 in order to identify groups. In order to take into account
the decrease of the magnitude range of the luminosity function sampled at increasing
distance, the galaxy luminosities were suitably increased with distance. 

The authors employed two variants of the P method to identify groups.
In the first variant the authors adopted a velocity 
link parameter of $cz = 200$ km/s and a distance link parameter of 
$0.31\;h^{-1}$ Mpc at the fiducial velocity of 500 km/s  
(the latter value corresponds to a number density threshold of 80). 
Both link parameters were then suitably scaled with distance. 
In the second variant the authors scaled with distance only the 
distance link parameter and kept the velocity link parameter constant 
at the value of 350 km/s. Given  the limited range of redshift encompassed 
by the NOG, this choice was used as an approximation to 
a slow scaling of $cz$ with distance, which was suggested by 
several cosmological N-body simulations (e.g., Nolthenius \& White 
1987; see Paper III for details on group identifications). 

Excluding the Local Group, in this paper we consider the 474 groups with at least
three members (of which 61 have at least ten members) obtained with the H method (H
groups) and the 506 groups with at least three members (of which 62 have at least
ten members) obtained with the second variant of the P method (P groups). About 45\%
of the NOG galaxies are found to be group members. For the sake of simplicity, in
this paper we do not consider groups obtained with the first variant of the P
method, because the two variants gave very similar catalogs of groups. 

The catalogs of NOG groups are among the largest catalogs of groups 
presently available in the literature. Being extracted from the same 
galaxy sample, the catalogs of H and P groups allow us also to investigate on 
differences in the group clustering properties related to differences 
in the algorithm of group identification. 

For the analyses presented in this paper, 
we construct volume-limited subsamples, which by definition 
contain objects that are luminous enough to be included in the sample when 
placed at the cutoff distance. 

Basically, we define volume-limited subsamples of galaxies with depth of 3000,  
4000, 5000, and 6000 km/s. They contain 1615, 1895, 2220, and 2257 galaxies, respectively. 
The absolute magnitude limits corresponding to these distances are $M_B=$-18.39, -19.01, -19.49, 
and -19.89 mag, respectively (here and throughout this paper, in general we omit 
the 5 $\log h$ term in the absolute magnitude).  

Moreover, we construct volume-limited samples of H and 
P groups with depth of 4000 km/s, by using suitably modified versions 
of the H and P algorithms in which the selection parameters which scale with distance are kept 
fixed at the values corresponding to 4000 km/s. These volume-limited 
samples contain 140 H and 141 P groups (with $n\geq 3$ members).

\section{Calculating the Two-Point Correlation Function} 

We define the separation in redshift space, $s$, as
 
\begin{equation}
s = H_{0}^{-1} \sqrt{V_i^2 + V_j^2 - 2 V_i V_j \cos\theta_{ij}}
\end{equation}

\noindent where $V_i$ and $V_j$ are the velocities of two objects (galaxies or 
groups) separated by an angle $\theta_{ij}$ on the sky. We calculate the  
CF in redshift space, $\xi(s)$, using the estimator proposed by Hamilton (1993)
 
\begin{equation}
\xi(s) = \frac{DD(s) RR(s)}{[DR(s)]^2} -1
\end{equation}

\noindent where $DD(s)$, $RR(s)$, and $DR(s)$ are the number of data--data,
random--random, and data--random pairs, respectively, with separation in the
interval $(s-ds/2, s+ds/2)$. Compared to the classical estimator of $\xi(s)$ by
Davis \& Peebles (1983), which explicitly depends on the mean density assigned to the
sample, Hamilton's estimator is less affected by the uncertainty in the mean
density, which is only a second--order effect. In particular, numerical simulations
have shown that Hamilton's estimator performs better on large scales where the
clustering is weak (see, e.g., Pons-Border\'ia et al. 1999 and Kerscher, Szapudi, \& 
Szalay 2000) for a thorough comparison of the reliability of several estimators 
of CF).  We generate the random
sample by filling the sample volume with a  random distribution of the same
number of points as in the data.  The random points are distributed in depth
according to the selection function of the data sample.

In order to minimize the statistical fluctuations in the determination of CF, 
we average the results obtained using many different replicas of the random sample. 
In general we compute 50 replicas in the case of galaxies and 400 in the case of groups. 

The counts $DD(s)$, $DR(s)$, and $RR(s)$ can be generalized to include a weight $w$,
which is particularly important to correct for selection effects at large distances
in a magnitude-limited samples. As a matter of fact, magnitude-limited samples
become sparser at larger distances due to the increasing loss of galaxies caused by
the apparent magnitude cutoff. This effect is quantified by the selection function
$S(s)$, which expresses the fraction of objects that are expected to satisfy the
sample's selection criteria. For $s<s^{*}$, where $s^{*}$ is a fiducial distance
that we take to be equal to $500\;km\;s^{-1}/H_0= 5\;h^{-1}$ Mpc, 
the appropriate selection function for
a magnitude-limited sample is simply $S(s)= f$, where $f$ is the average
completeness level of the sample ($f =0.97$ for the NOG), whereas for $s>s^{*}$ 
the selection function is 

\begin{equation}
S(s) = f \frac{\int_{max[L_s, L_{min(s)}]}^{\infty} \Phi(L) dL}{\int_{L_s}^{\infty} 
\Phi(L) dL}
\end{equation}

\noindent where $\Phi$ is the luminosity function, $L_{min}(s)$ is the minimum luminosity
necessary for a galaxy at a distance $s$ (in Mpc) to make it into the sample and the
integral is cutoff at the lower limit of $L_s=L_{min}(s^{*})$. $L_{min}(s)$
corresponds to the absolute magnitude $M_B = - 5 \log s -25 + B_{lim}$, where
$B_{lim}=14$ mag is the limiting apparent magnitude of NOG; thus $L_s$ corresponds
to the absolute magnitude $M_s = -14.50$ mag. 
 
In the calculation of the selection function 
we use the Schechter (1976) form of the luminosity function that we derive 
by means of Turner's (1979) method (see also de Lapparent, Geller, \& Huchra 1989,  
and Paper II), using redshifts as distance indicators, for the whole sample 
and for specific morphological subsamples. The adopted Schechter parameters are 
$\alpha=-1.2$ and $M_B^*=-20.0$ (see Paper III) for the whole NOG (see also paper II 
for the values appropriate for the specific morphological types).  We have checked 
that our results on CFs are insensitive to changes of $M^{*}$ 
and $\alpha$ by 0.2 mag and 0.1, respectively. 
 
Unless otherwise specified, we compute the weighted CF by
replacing the counts of pairs with the weighted sum of pairs, $\sum w_i w_j$, which
takes into account the selection effects acting on the sample used. The weighting
scheme we adopt is that of equally weighted volumes, $w_i=1/S(s_i)$, where $S(s)$
is the selection function. We do not use the minimum variance weighting scheme
(Davis \& Huchra 1982), which requires prior knowledge of the CF. 
However, for a dense sample like the NOG, the practical improvement provided by
using the minimum variance method is very small (Hamilton 1993; Park et al. 1994). 

In several cases we use volume-limited subsamples.  
This is equivalent to applying a lower limit to 
the luminosity of the objects of the subsample, leading to an uniformly 
selected data set with the same weight assigned to each object ($S(s)=f,\; 0\leq s\leq 
s_{max}$).   
  
We calculate the errors for the  $\xi(s)$ by using 100 
bootstrap resamplings of the data (e.g., Ling, Frenk, \& Barrow 1986);   
bootstrapping tends to overestimate the real errors, as shown by Fisher et 
al. (1994).

Since in general $\xi(s)$ is satisfactorily described by a power law over 
a fairly large interval of $s$, we always fit $\xi(s)$ to the form 
$\xi(s)=(s/s_0)^{-\gamma}$ with a non-linear 
weighted least-squares method in the intervals where $\xi(s)$ is reasonably 
fitted by a single power-law. Since the points are not statistically independent, 
it is not strictly correct to use a least-squares method for doing the fit. 
But it is a satisfactory approximation (Bouchet et al. 1993) especially if we bear 
in mind that in any case the results of the power-law fits are sensitive to the 
interval adopted in the fits. Besides, the underestimation of the fit errors 
due to the bin-bin interdipendence is well compensated by the overestimation 
given by the bootstrapping-resampling method (Mo, Jing, \& B\"orner 1992). 

The variance of galaxy counts in the volume $V$ is related to the 
moment of $\xi(s)$ which gives the separation--averaged value of 
$\xi(s)$ in the volume $V$, 
\begin{equation} 
\sigma^2 = \frac{1}{V^2} \int_{v} dV_1 dV_2 \xi(|(\vec{s_1} - \vec{s_2}|) 
\end{equation} 
For a power-law correlation function $\xi(s) = (s/s_0)^{-\gamma}$ and a spherical 
radius of radius $R$, this integral can be performed analytically to yield 
(cf. Peebles 1980, section 59.3)

\begin{equation}
\sigma^{2}(R) = \frac{72 (s_0/R)^{\gamma}}{2^{\gamma} (3 - \gamma) (4 - \gamma) (6
- \gamma)}
\end{equation}

We use this expression to calculate the {\it rms} fluctuation in galaxy counts within a
sphere of 8 $h^{-1}$ Mpc radius, $\sigma_8$, a quantity which is often used to normalize
theoretical models. 

We evaluate the relative bias between two samples of galaxies of different 
luminosities or morphological types through the expression 

\begin{equation}
\frac{b}{b_*}(s) = \sqrt{\frac{\xi(s)}{\xi_*(s)}}
\end{equation}

\noindent where the starred symbols denote a sample taken as a fiducial.

\section{The correlation function of NOG galaxies}

\subsection{Results for the whole NOG}
The redshift-space weighted CF from the whole NOG sample, (7028 galaxies 
in the velocity range $50< cz
\leq 6000$ km/s) is satisfactorily described by a power law in the interval $1
\lesssim s \lesssim 15 h^{-1}$ Mpc.  Error bars become progressively larger beyond
$\sim10 h^{-1}$ Mpc.

From a power-law fit calculated in the interval $2.7 - 12 h^{-1}$ Mpc, which in this
case and in most subsamples of galaxies examined in this paper is the one where
$\xi(s)$ is best-fitted by a single power law, we obtain a correlation length of
$s_0 = 6.42\pm0.11\;h^{-1}$Mpc and a slope $\gamma = 1.46\pm0.05$.  From these
values we estimate a value of $\sigma_{8}=1.03\pm0.02$.

More specifically, $\xi(s)$ is not a perfect power-law. It tends to flatten on small scales 
($s<2\;h^{-1}$ Mpc) because of the effects of peculiar motions, which 
smear out galaxy systems along the radial line-of-sight. It  
tends to steepen on large scales ($s>15\;h^{-1}$ Mpc). For
instance, we find $s_0=6.12\pm0.10\;h^{-1}$ Mpc, $\gamma = 1.39\pm0.03$ 
in the range $1.1\leq s \leq 17\;h^{-1}$ Mpc, and $s_0= 6.24\pm0.09\;h^{-1}$ Mpc, 
$\gamma= 1.58\pm0.05$ in the range $2.7\leq s \leq 17\;h^{-1}$ Mpc. 

Our values of $s_0$ and $\gamma$ turn out to be in good agreement with those
relative to the redshift-space $\xi(s)$ derived from the SSRS1 sample 
(Maurogordato, Schaeffer, \& da Costa 1992; $s_0\sim 5 - 8.5\;h^{-1}$ Mpc, 
$\gamma\sim1.6$), SSRS2 sample (Willmer et al. 1998; $s_0=5.85\pm0.13\;h^{-1}$Mpc, 
$\gamma=1.60\pm0.08$), the sparsely
sampled Stromlo--APM survey (Loveday et al. 1995; $s_0=5.9\pm0.3\;h^{-1}$ Mpc,
$\gamma=1.47\pm0.12$), the Las Campanas Redshift Survey (LCRS; Tucker et al.
1997; $s_0=6.28\pm0.27\;h^{-1}$Mpc; $\gamma=1.52\pm0.03$).

Besides, the amplitude of the NOG $\xi(s)$ is consistent with that of the
Optical Redshift Survey (ORS; Hermit et al. 1996; $s_0=6.6 -6.8\;h^{-1}$Mpc, $\gamma=1.57,
1.51$ for the ORSd and ORSm samples, respectively) and that of the CfA2 survey
(de Lapparent, Geller, \& Huchra 1988) ($s_0=7.5\pm1.7\;h^{-1}$Mpc,$\gamma=1.6\pm0.1$), which 
encompasses the Great Wall, an overdense region characterized
by a very high degree of local clustering ($s_0\sim 15\;h^{-1}$ Mpc, see 
Ramella, Geller, \& Huchra 1992). 

In Figure 1 we compare the redshift-space galaxy CF from the 
whole NOG with the redshift-space galaxy CFs  
from some recent redshift surveys, i.e. the Stromlo-APM (Loveday et 
al. 1995), LCRS (Tucker et al. 1997), Durham-UKST (Ratcliffe et al. 1996), 
and EPS (Guzzo et al. 2000) redshift surveys. 

Our results are in good agreement with the results from Stromlo-APM and especially
with those from LCRS, although the volumes of the last two surveys ($2.5\cdot
10^6\;h^{-3}$ Mpc$^3$ and $2.6\cdot 10^6\;h^{-3}$ Mpc$^3$, respectively) are quite
larger than that of NOG ($5.95\cdot 10^5\;h^{-3}$ Mpc$^3$).  On small scales
($s<6\;h^{-1}$ Mpc), our results also agree with those from the sparsely-sampled
Durham-UKST survey (encompassing a volume of $4\cdot 10^6\;h^{-3}$ Mpc$^3$), which,
however, around $s\sim10\;h^{-1}$ Mpc, leads to a $\xi(s)$ characterized by a
somewhat large amplitude. On the other hand, the amplitude of the NOG $\xi(s)$ is
slightly larger than that of the ESP sample (characterized by $s_0\sim 5\;h^{-1}$
Mpc, $\gamma\sim 1.5$), which, owing to its faint magnitude limit, probably contains
a much larger population of intrinsically faint, less clustered galaxies. 

Noticeably, we obtain a smoother $\xi(s)$ than do several previous works probing
larger volumes (see, e.g. the results from the Stromlo-APM and Durham-UKST surveys
plotted in Figure 1) because NOG contains a large number of galaxies. 

\subsection{Results for different subsamples}
An important issue to test is the reliability of the weights assigned to each
galaxy. We do that by comparing magnitude-limited subsamples 
truncated  at a given maximum distance
$cz_{max}$ to their volume-limited counterparts. 
The magnitude- and volume-limited subsamples probe
the same volume in space and therefore should lead to the same CF, if the
weight in the magnitude-limited subsamples are unbiased, unless there is a strong
dependence of clustering on luminosity. 

Results of this test are presented in Figures 2 and 3. Figure 2 shows the weighted and
unweighted  CFs  of the whole NOG sample
(7028 galaxies) and the volume-limited sample limited at 6000 km/s (2257 galaxies). 
Analogously, Figure 3 shows the comparison
between the CF of the magnitude-limited (4364 galaxies) and
volume-limited (1895 galaxies) subsamples limited at 4000 km/s.

In Figures 2 and 3 the CFs of the volume-limited samples satisfactorily agree with
the weighted CFs of the magnitude-limited samples. This reassures of the 
validity of the selection function adopted for the NOG.
On the other hand, the unweighted CFs have markedly 
smaller amplitudes. The differences between the results of the two
weighting schemes (in this case as well as in other cases mentioned below) stem
from the fact that the weighted CF weights each volume of space equally and
therefore traces better the clustering of more distant and luminous objects, whereas
the unweighted CF is more sensitive to the clustering of nearer and fainter
objects. In general, throughout this paper we find that the amplitudes of the weighted
CFs are significantly larger than those of their unweighted counterparts and
that magnitude-limited and volume-limited subsamples restricted to smaller depths 
have CFs of smaller amplitudes with respect to subsamples of greater depths.
(cf. the CFs of Figures 2 and 3). The latter effects are apparent also 
in several previous works (e.g. Hermit et al. 1996; Willmer et al. 1998).  
 
Both pronounced tendencies are at least partially due to 
an increasing clustering with luminosity (we discuss this
issue in \S 6). This effect also explains why indeed the volume-limited sample with 
depth of 6000 km/s tends to lead to a 
slightly greater CF than the corresponding magnitude-limited sample (see 
Figure 2).

There are appreciable variations in CFs among small subsamples of NOG. 
For instance, we have checked that the CF of the subsample (out to 6000 km/s) 
relative to the strip of the sky between the 
UGC and ESO galaxy catalogs ($-17.5^{\circ} \leq \delta \leq -2.5^{\circ}$) 
is characterized by an appreciably smaller amplitude and a steeper 
slope than that of the whole sample, from which the CFs of the regions covered 
by UGC and ESO (i.e. $\delta> -2.5^{\circ}$ and $\delta< -17.5^{\circ}$, respectively)
do not appreciably depart. Thus, we confirm the results by Hermit et al. (1996), 
who, relying on the diameter-limited ORS sample, noted the  peculiar clustering 
properties of the above-mentioned strip in the nearby universe.

\section{Morphological Segregation}

We analyze the dependence of clustering on galaxy morphology, calculating the
CFs for different morphological types. Studies of morphological segregation 
based on the comparison of two-point correlation functions are complementary to analyses 
which deal with the relation between galaxy morphology and local galaxy number density, 
the morphology--density relation.  This relation was defined by Dressler (1980) for 
galaxy clusters and most of relevant analyses refer to cluster regions, i.e., 
to scales smaller than $\sim 1.5\;h^{-1}$ Mpc (see, e.g., Andreon, Davoust, \& 
Heim 1997 and Dressler et al. 1997 for recent accounts of the morphology-density 
relation). But it was soon  suggested that this relation extends outside clusters, 
i.e. in groups (e.g., Bhavsar 1981, de Souza et al. 1983) and in the field (Postman \& 
Geller 1984). Investigations on morphological segregation have also exploited 
other statistical descriptors of galaxy clustering (e.g., Lahav \& Saslaw 
1992; Dom\'inguez-Tenreiro et al. 1994).

Figure 4 shows the weighted CFs for
the E-S0 and Sp objects of the whole sample and the magnitude-limited sample out to
4000 km/s. For these magnitude-limited samples and for all volume-limited samples
considered, E-S0 galaxies are always more clustered than spirals out to large scales
($\sim20\;h^{-1}$ Mpc). Thus we confirm the results of many previous works on the
morphological segregation between early- and late-type galaxies   
(e.g., the early works by Davis \& Geller 1976; Giovanelli, Haynes, \& Chincarini 1986;  
Iovino et al. 1993). 

Remarkably, analyzing the volume-limited sample with depth of 3000 km/s, we find
that an appreciable morphological segregation between early- and late-type galaxies
is present also at relatively low luminosities (i.e., at $M_B>-19$ mag), in spite of
some recent doubts expressed by Beisbart \& Kerscher (2000) who claimed that this
morphological segregation concerns essentially luminous galaxies. Thus, our 
results support the contention (e.g., the review by Ferguson \& Binggeli 1994) 
that the well-known morphology-density relation inferred for the classical Hubble 
morphological types (Dressler 1980) holds also for dwarf galaxies. 

We also attempt to use a fine morphological subdivision. 
Figures 5 and 6 show the weighted CFs for 6 morphological 
types, E, S0, S0/a, Sa, Sb, and Scd (see \S 2 for details on the 
adopted morphological subdivision) for the whole NOG sample.

Table 1 reports the power-law 
fit parameters for the various morphological types . These calculations 
refer to the interval of separations (mostly the 
$2.7 \leq s \leq 12\;h^{-1}$Mpc range)  
where the CF is in general well approximated by a power law.

If we had attempted to subdivide the E galaxies into two subtypes 
(i.e., $T<-3.5$ and $-3.5\leq T< -2.5$) we would have found two 
indistinguishable CFs for the appropriate selection functions.  
Analogously, if we had subdivided the Sa galaxies into two subtypes 
(i.e., $0.5\leq T <1.5$ and $1.5\leq T <2.5$) we would have obtained 
two indistinguishable CFs for the appropriate selection functions. 

Using the whole magnitude-limited sample is an attempt to extract 
the maximum signal from the available data, but can have some 
drawbacks in the presence of luminosity segregation (see \S 6), if 
the type-specific luminosity functions are not similar. In particular, it is 
known that Scd galaxies are typically less luminous than the other types 
(see, e.g., Paper II). Therefore, we have considered also the 
morphological subsamples extracted from the volume-limited sample 
at 6000 km/s. Table 2 reports the power-law fit parameters for these 
subsamples, except for the 101 S0/a galaxies, whose CF is very noisy 
and can not be reasonably approximated by a power law. 

Figures 5 and 6 and Tables 1 and 2 show that, in general, also 
morphological subtypes exhibit an appreciable morphological segregation. 
In general, especially at intermediate scales (around $5\;h^{-1}$Mpc)  
there appears to be a gradual decreasing of the strength of clustering 
from the early-types, specifically from S0 galaxies, to the latest 
types, the Scd galaxies. In several cases the morphological segregation 
between the various types persists out to large scales ($s\sim15\;h^{-1}$Mpc). 
As regards the E galaxies, these objects do not exhibit a greater clustering than 
the S0s. Remarkably, if we had subdivided the E-S0 objects into earlier and later types  
choosing a different limit in T, i.e. $T<-3.5$ and $T\geq -3.5$,  
we would have found again that the earlier types do not cluster more 
strongly than the later ones.

The relative differences   
in the clustering of the various morphological types are confirmed 
by the analysis of other magnitude-limited samples  
and volume-limited samples (e.g., the samples truncated at 4000 km/s). 
Besides, our results on $s_0$ do not change much if we calculate 
power-law fits keeping $\gamma$ fixed at the value of 1.5, which is 
typical for the CF of the whole NOG.

A comparison between Table 1 and 2 reveals that, for all morphological 
subtypes,  the CFs derived from volume-limited samples tend to have greater  
amplitudes than those derived from the corresponding magnitude-limited 
samples, which is likely to be an effect of luminosity 
segregation (see \S 6). Besides, at least partly for the same reason, 
morphological subsamples limited to smaller depths tend to have CFs of 
lower amplitudes, as in the general case (see \S 4.2).

Although the statistical significance of the morphological segregation 
can be based on the values of the fit parameters (together with 
associated errors) reported in Tables 1 and 2, it is useful to 
provide a further check on this effect. For example, we have 
decided to test the significance of the difference between the 
CF amplitudes of the volume-limited samples (at the depth of 6000 km/s) 
of the E-S0 and Sp objects, whose $s_0$-values are different at the 
$\sim6 \sigma$ confidence level. We have addressed this point by randomly selecting 
two sets of objects of sizes equal to the sizes of the two volume-limited 
samples from wich they are extracted. We have compared 
the two resulting CFs, which are typically similar. For 500 random selections, 
the mean ratios between the CF amplitudes are always smaller than the observed value 
(2.5 in the usual $s$-range) for the two volume-limited samples.
This implies that the two CFs are different at the $>$99.8\% confidence 
level.   

Our fitting $s_0$-value for the E-S0 objects better agrees with the 
high values of $s_0$ reported for the early-type galaxies of the   
Stromlo-APM and Perseus--Pisces samples (i.e. $s_0=9.6\pm0.2\;h^{-1}$Mpc in the 
redshift space and $s_0=8.4\pm0.8\;h^{-1}$Mpc in the real space; see Loveday et 
al. 1995 and Guzzo et al. 1997, respectively) 
than with the relatively low values of $s_0$ 
($s_0\sim6-7\;h^{-1}$Mpc) reported for the early-type objects of the SSRS1, 
(Santiago \& Da Costa 1990), ORS (Hermit et al. 1996) and SSRS2 
(Willmer et al. 1998) surveys.

On the other hand, our $s_0$-value for Sp galaxies appears to be in substantial 
agreement with several relevant previous results 
($s_0 = 5.3 - 5.4\;h^{-1}$ Mpc according to Loveday et al. 1995 and Willmer et al. 1998), 
whereas it is somewhat greater than the value ($s_0\sim 4.5\;h^{-1}\;Mpc$) 
reported by Santiago \& Da Costa (1990) for the SSRS1 sample.
Hermit et al. (1996) subdivided the ORS spirals into Sab and Scd types.
The $s_0$-values  reported by the authors for these two types are in substantial 
agreement with the values we obtain for the Sb and the Scd types, respectively. 

The lower amplitude of the CF of the IRAS 1.2 Jy redshift survey (Fisher et
al. 1994) ($s_0=4.53\pm0.22\;h^{-1}$Mpc, $\gamma=1.28\pm0.04$) compared with that of
NOG and many other optical samples, for all galaxies and for all spirals, reflects
the relative bias that exists between optical- and infrared-selected galaxies. The
CF amplitude of the NOG Scd galaxies approaches that of the IRAS samples 
(see also Seaborne et al. 1999 and Szapudi et al. 2000 for the CF of the PSCz sample). 

Figure 7 shows the relative bias ($\sim 1.7$) between the E-S0 and Sp objects (see eq. 6)  
of the whole NOG as a function of the scale for the whole NOG. This bias appears to be 
substantially constant with scale. The relative bias ($\sim 1.8$) between the E-S0 
and Scd objects turns out to be  also constant with scale. 
These results are confirmed by the inspection of magnitude-limited and volume-limited samples 
with smaller depths. We have verified that also the relative biases between 
other morphological types appear to be substantially constant with scale at least 
up to $\sim10\;h^{-1}$Mpc. As a matter of fact, all fitting values of $\gamma$ 
relative to the various morphological types appear to be substantially 
not inconsistent with the value of $\gamma\sim1.5$ which pertains to the whole 
NOG sample. In other words, there is no sure evidence for an increasing or 
decreasing value of $\gamma$ as we go from early to late types. 

Our results disagree with those by Willmer et al. 
1998 (confirmed by the reanalysis of Beisbart \& Kerscher 2000), who found 
that the relative bias between the SSRS2 E-S0 and Sp (calculated from 
both redshift-space and real-space correlation functions) decreases from 
small ($\sim1\;h^{-1}$Mpc) to relatively large ($s\sim8\;h^{-1}$Mpc) scales.
Also Guzzo et al. (1997) reported a similar scale-dependent bias in their 
real-space correlation function analysis of the Perseus--Pisces region.
Hermit et al. (1996) claimed a similar effect (in redshift space) for the  
early-type and late-type ORS objects in the $1 - 10\;h^{-1}$ Mpc range . But 
the bias appears to grow with scale from $\sim10\;h^{-1}$Mpc to $\sim20\;h^{-1}$Mpc 
so that on large scales ($\sim 20\;h^{-1}$Mpc) it is not substantially different from 
the value relative to small scales ($\sim 1\;h^{-1}$Mpc). 
Furthermore, there is no clear evidence for a significant scale-dependence 
of the bias (evaluated in redshift space) in the Stromlo-APM sample (Loveday et al. 1995) 
as well as in the SSRS1 (Santiago \& da Costa 1990) survey, but the real-space analysis 
of the former sample hints at some effect. 

In general, the morphological segregation can not be caused by a possible 
luminosity segregation (with luminous galaxies clustering stronger than 
dim galaxies). As a matter of fact, the shape of the luminosity function of 
the E-S0 galaxies does not differ appreciably from that of Sp galaxies.  
Moreover, there are no large differences between the 
shapes of the type-specific luminosity functions, except for the 
earliest (E) and latest (Sm-Im) types (see Paper II).

\section{Luminosity Segregation}

Several investigations concerned with luminosity segregation looked particularly at a 
sequence of volume-limited samples and compared the amplitudes of the corresponding 
CFs. An increasing amplitude of CF with growing depth of the sample 
is in general considered as an indication of luminosity segregation (with luminous 
galaxies having a stronger clustering than dim objects). We have verified that 
this tendency is clearly apparent also in the NOG volume-limited samples. However, 
the results can be affected by the fact that one compares the clustering of 
low-luminosity and high-luminosity galaxies in very different volumes (the former 
objects are necessarily closer to us than the latter ones). The rise of the CF  
amplitude with the depth of the sample could be explained even in terms of a 
fractal model for the galaxy distribution, with no luminosity segregation, as 
shown by Pietronero (1987). 

Therefore, we have decided to look for luminosity segregation by comparing the 
clustering of galaxies of different luminosities within the same volume-limited sample.
 
First we consider the volume-limited sample with depth of 6000 km/s. Figure 8 
presents the CFs for the objects of different luminosity classes (-19.89 is the
limiting value of $M_B$ of this volume-limited sample). Clearly, the CF amplitude 
decreases with decreasing luminosity, indicating a significant luminosity
segregation.  Table 3 shows the parameters resulting from the corresponding
power-law fits together with the number density of galaxies $n_g$ (for a 
completeness of 97\%) , and the mean intergalaxy distance $d$, calculated as 
$d= n_g^{-1/3}$. 

Remarkably, the values of $s_0$ reported in Table 3 are significantly 
greater than the values predicted from the $s_0$ - $d$ relationship, 
$s_0=0.4 d$, proposed by Bahcall \& West (1992) for a variety of galaxy 
systems. The departure from this relationships gets smaller for 
brighter galaxies, in agreement with Cappi et al.'s (1998) finding 
based on the SSSRS2 survey.  

Figure 9 reveals that the relative biases between the above-mentioned subsamples of
luminous galaxies and the least luminous ones is $\sim$1.5, 1.3 and 1.1,
respectively. Clearly, these relative biases appear to be substantially constant
with scale at least up to $10\;h^{-1}$Mpc, in agreement with the contentions of
Benoist et al. (1996) and Willmer et al. (1998). Consistently, a power spectrum 
analysis of volume-limited samples drawn from the CfA2 survey revealed  
a luminosity segregation with a luminosity bias independent of scale 
(Park et al. 1994).

As regards the volume-limited sample with depth of 5000 km/s, 
we subdivide the objects of this sample into three luminosity intervals, 
$M_B\leq -20.3$ ($N=632$), $M_B\leq -19.89$ ($N=1267$) and $M_B\leq -19.49$ ($N=2220$) 
(the last value is the limiting $M_B$ of the sample). Also in this sample we 
detect an appreciable luminosity segregation (see the plot of the 
corresponding CFs in Figure 10), with relative biases which are confirmed 
to be substantially constant with scale at least up to $10\;h^{-1}$Mpc.

Analogously, we subdivide the objects of the volume-limited sample with depth of
4000 km/s into three luminosity intervals, $M_B\leq -19.89$ ($N= 624$), $M_B\leq
-19.49$, ($N=1107$), $M_B\leq -19.01$ ($N= 1895$), and the objects of the
volume-limited sample with depth of 3000 km/s into four luminosity intervals, the
three mentioned below (comprising $N=$ 325, 556, and 953 galaxies, respectively) and
the interval $M_B\leq -18.39$.  ($N=1615$). The last $M_B$-intervals considered
reach the limiting $M_B$-values of the samples. 

We find no evidence of luminosity segregation below a luminosity corresponding to 
$M_B=-19.49$ mag, since the CFs relative to the low-luminosity intervals do not 
differ appreciably. We have checked that the removal of the galaxies belonging 
to the region of the Virgo cluster, which is the most prominent {\it finger of God}  
in the nearby ($cz<3000$ km/s) universe, does not change this conclusion. 

Repeating the same kind of analysis of the volume-limited samples for the E-S0 and
the Sp objects, separately, we verify that the luminosity segregation holds
separately for early- and late-type luminous ($M_B\leq -19.49$ mag) galaxies, whereas
it is not present in the dim objects of both types.  In conclusion, both early- and
late-type galaxies show luminosity segregation, which, however, starts to appear
only for luminous galaxies ($M_B\lesssim -19.5$ mag). 

As in the case of morphological segregation (see \S 5), we wish to  provide a 
further test of the luminosity segregation, considering specifically  
two volume-limited samples at the depth of 6000 km/s, the samples of galaxies 
having $M_B\leq -19.89$ mag and $M_B\leq -20.6$ mag. According to the power-law fits 
reported in Table 3, the difference between the correponding $s_0$-values is 
significant at the $\sim 5 \sigma$ confidence level. We randomly select 499 galaxies 
from the former sample and compare its CF with that of the former sample. The two 
CFs are in general similar and, for 500 random selections, we can establish 
that the observed difference in the CF amplitudes of the two volume-limited 
samples (the mean ratio is 1.4 in the usual $s$-range) is significant at 
the 99.8\% confidence level.

Our results are in line with a series of papers which, especially in the recent 
literature, reported evidence of diameter (Maurogordato et al. 1992) and luminosity 
segregation (e.g., Hamilton 1988, Davis et al. 1988, B\"orner, 
Mo, \& Zhou 1989, Dom\'inguez-Tenreiro \& Mart\'inez 1989, Park et al. 1994,  
Guzzo et al. 1997, Willmer et al. 1998; Benoist et al. 1996; see the last paper 
for a list of articles arguing against luminosity segregation).

But in the literature there is less consensus about 
the range of luminosities and the morphological types at which the effect occurs. 
For instance, to cite some earlier results which disagree with our findings, 
Loveday et al. (1995) claimed to detect luminosity segregation only 
for galaxies of relatively low luminosity within the Stromlo-APM 
sample. Specifically, the authors found no evidence for stronger clustering of 
objects with $L>L^{*}$ ($-22\leq M_{bj} \leq -20$) compared to galaxies of 
intermediate luminosities ($-20\leq M_{bj} \leq -19$, $L\sim L^{*}$) (except 
for a small effect observed for late-type galaxies only), whereas the 
latter objects were found to be more clustered than dim galaxies 
($-19 \leq M_{bj} \leq -15$, $L<L^{*}$).

Hasegawa \& Umemura (1993) claimed that, after correction of luminosities for
internal extinction, the early- and late-type galaxies of the CfA1 survey show
luminosity segregation of opposite sign (i.e. segregation and anti-segregation,
respectively) so that this opposite behavior is responsible for a lack of luminosity
segregation in the total sample.  Beisbart \& Kerscher (2000), who reanalyzed the
SSRS2 data in the mathematical framework of marked point processes, found a
scale-dependent luminosity segregation attributed mostly to early-type objects. 

Interestingly, no luminosity segregation was found in the IRAS 1.2 Jy sample 
(Bouchet et al. 1993, Fisher et al. 1994) and in the PSCz sample  
(Szapudi et al. 2000). A possible explanation for the different behavior of 
optical and IRAS galaxy samples with respect to luminosity is that the optical 
magnitudes are likely to be more strongly related to the mass than far-infrared fluxes.

The dissimilarity between optical and IRAS samples of galaxies (and spirals) 
can be evidenced by the fact that only 53\% of NOG galaxies are common to the PSCz;  
moreover, 30\% of the PSCz objects contained in the volume encompassed by the NOG 
has no counterparts in the NOG.

Lets us inspect the sample of very luminous galaxies (hereafter VLG), 
i.e. the objects with $M_B\leq-21$ mag 
($\grsim 2.4\;L^*$) (see Table 3). The morphological 
mix of our VLG sample differs from that of the whole NOG sample, because the 
former sample comprises a larger fraction of E-S0 galaxies (26\% versus 15\%) 
and a smaller proportion of Scd galaxies (39\% versus 53\%). Compared to the whole 
NOG sample, there are fewer VLG objects among field (ungrouped) galaxies.  
For instance, for the H groups, 
the fraction of VLG objects which are field galaxies,  
members of pairs, members of groups with at least 3, 10, and 20 members 
is 18\%, 22\%, 60\%, 26\%, and 15\%, respectively (for the whole NOG, 
the corresponding values are 39\%, 17\%, 44\%, 20\%, and 13\%). 
Moreover, we count 21 (10\%) VLG objects among members of the poor and rich 
clusters identified within the NOG (i.e., A194, A262, A569, A3229, 
A3565, A3574, A3656, A3742, Centaurus, Dorado, Eridanus, 
Fornax, Hydra, Ursa Maior, Pegasus, Virgo; see Table 5 of Paper III); these 
systems comprise 10\% of NOG galaxies. The three richest clusters 
(richness class 1) in the NOG volume (Virgo, Hydra, A3565) 
comprise 6 (3\%) VLG objects and 5\% of 
NOG galaxies. Finally, if we count the number of VLG objects and the total 
number of galaxies belonging to the few systems 
which have the largest velocity dispersions (e.g. $\sigma_v>400$ km/s) 
and virial masses (e.g., $M_v>10^{14}\;h^{-1}\;M_{\odot}$), we find again 
similar percentages (see \S 8.1 and \S 8.3 for the calculation of the 
two above-mentioned quantities).

Hence, VLG reside preferentially in galaxy systems (pairs and groups), 
though not predominantly in  rich systems like galaxy clusters.   
Consistently, they show a value of $s_0$ (see Table 3) which is likely  
to be consistent with that of fairly rich groups, but it 
is still lower, though not by much, than that of clusters. We have checked 
that, if we define the VLG as the objects with $M_B$ brighter than values 
ranging from -21.1 to -21.5 mag ($\sim 3.8 L^{*}$), we obtain CFs 
of similar amplitudes. 

Our results on VLG are in partial agreement with those 
by Cappi et al. 1998, who, examining the VLG objects (with $M_{bj}\leq$-21 mag, 
$L\grsim 4 L_{bj}^{*}$) extracted from the
SSRS2 redshift survey, stated that these objects are not preferentially 
located in bona fide clusters, but stressed the similarity between their 
correlation length (for which they obtained $s_0= 16\pm 2\;h^{-1}$ Mpc) 
and that of APM clusters of low richness (for which $s_0\sim 14\;h^{-1}$ Mpc 
according to Dalton et al. 1994 and Croft et al. 1997).

\section{The correlation function of NOG groups}

In general the redshift distributions of relatively rich P and H groups are shifted to
smaller values than that of galaxies. However, we have verified that the redshift
distributions of the 506 P groups with $n\geq3$ members, the 474 H groups with $n\geq$ 3,
the 280 and 189 H groups with $n\geq$ 4 and 5 members, respectively, 
are not significantly different from that of galaxies, according to the nonparametric
Kolmogorov--Smirnov (KS) statistical test (e.g., Hoel 1971). Therefore, we have computed the
CF for these groups, assuming the same selection function adopted for galaxies. 
The same assumption was made by Ramella et al. 1990, Trasarti-Battistoni et al. 1997, 
and Girardi et al. 2000. 

In Figure 11 we show the weighted CFs for P groups (with $n\geq$3), H groups
(with $n\geq$ 3 and 5) and all galaxies.  On small scales ($<3.5\;h^{-1}$ Mpc) the
CF of groups starts dropping because of the anti-correlation due to the
typical size of groups, whereas on large scales ($>20\;h^{-1}$ Mpc) the
signal-to-noise ratio of CF appreciably decreases.  Thus, we limit our
following analysis to the separation interval $3.5\leq s \leq 20\;h^{-1}$ Mpc. The
CFs of groups have larger amplitudes than that of galaxies, especially for the
H groups with $n\geq 4$ and $n\geq 5$ members.  Over the above-mentioned interval of
separations, the mean value $<r>$ of the ratios (11 values) between the CF of
groups and that of galaxies turns out to be $1.52\pm0.32$ and $1.97\pm0.80$ for the
P and H groups with $n\geq$ 3, respectively, and somewhat larger, i.e.
$2.40\pm0.70$, $2.36\pm0.54$, for the H groups with $n\geq$ 4 and 5, respectively.
These numbers do not change much, if we calculate the ratios using a different
number of points over a similar interval of separations. According to these results,
the excess of clustering of groups with respect to that of galaxies is significant
at the $\sim2 \sigma$ confidence level. 

Our results appear to be in good agreement with those by Girardi
et al. (2000), who found $<r>=1.6$ for their total and volume-limited 
samples of percolation groups with $n\geq$3 members. Our results are also compatible with 
those by Trasarti-Battistoni et al. (1997), who found $<r>\sim 2$ for their 
total sample of percolation groups with $n\geq$3 members selected in the Perseus--Pisces region 
above a similar number density threshold (110).  
These authors noted that the degree of clustering of percolation groups 
is insensitive to the choice of the velocity link parameter, but is a little 
sensitive to the distance link parameter adopted (values denoting  
much greater density contrast yield a stronger clustering). 

The unweighted CFs of all groups and galaxies are
smaller than the corresponding weighted CFs and are equal within 1 $\sigma$
errors. Hence distant groups, which are, on average, more luminous and more 
massive, are more strongly correlated than nearby groups.

Exploiting the volume-limited samples of groups and galaxies (at the depth of 4000
km/s), we can extend this kind of analysis to the relatively rich groups.  In 
Figure 12 we show the CFs for the P groups with $n\geq$ 3, 5 members and for 
galaxies. Groups (especially those 
with $n\geq 5$ members) again tend to show a stronger clustering than galaxies, 
although, because of the large errors involved, the effect is not very 
significant. Table 4 presents the mean value of the ratio $<r>$ calculated for the 
above-mentioned interval of separation. 

In order to evaluate the significance of the difference in clustering between
groups and galaxies irrespective of the number of bins in which we divide a  
given interval of separations, we have rebinned the pair counts
considering the wide interval $0.5 \leq \log s \leq 1.1\;h^{-1}$ Mpc. Table 5 lists
the amplitudes of the $\xi(s)$ (together with the $1\sigma$ errors)
calculated at the mean value of the above-mentioned interval, $\log s^{*} = 0.8$
(i.e. $s^{*} = 6.31\;h^{-1}\;$Mpc) for the magnitude-limited (ml) samples and 
volume-limited (vl) samples of groups and galaxies mentioned above.  
This table confirms an excess of clustering
of groups with respect to galaxies at the $~1 - 2\;\sigma$ confidence level,
especially for groups with $n\geq$ 4 members.

Using a power-law fit over the interval $3.5 \leq s \leq 20\;h^{-1}$ Mpc
for the weighted CF of the total sample of groups, we obtain 
$s_0= 8.4\pm0.7\;h^{-1}$ Mpc ($7.8\pm0.4\;h^{-1}$ Mpc), $\gamma=1.3\pm0.2$ 
($2.0\pm0.2$) for the total sample of H groups (P groups) with $n\geq$ 3 members,  
$s_0=9.4\pm1.0\;h^{-1}$ Mpc, $\gamma=1.6\pm0.3$ for the H groups with $n\geq$4 members, 
$s_0=9.8\pm1.2\;h^{-1}$ Mpc, $\gamma=1.7\pm0.4$ for the H groups with $n\geq$5 members. 
The corresponding values of $\sigma_8$ are $1.2\pm0.1$ ($1.4\pm0.2$), $1.4\pm0.2$, 
$1.6\pm0.3$, respectively. 

Given the formal errors and the variations among results coming from different 
group samples, we can conclude that the slope of the CF of groups is not inconsistent 
with that of galaxies (the above-mentioned results point to a  mean value of $\gamma\sim1.6$), 
whereas $s_0$ is clearly greater 
than that of galaxies (at nearly the $3 \sigma$ confidence level) and ranges from $s_0\sim8\; 
h^{-1}$ Mpc (for groups with $n\geq 3$) to  $s_0\sim 10\;h^{-1}$Mpc (for $n\geq 5$). 

The strength of clustering of the P and H groups does not differ significantly. 

Our results agree with those reported by Girardi et al. (2000) for their 
volume-limited sample of groups at the depth of 7800 km/s. For this sample, which 
has a median velocity dispersion of $\sigma_v = 214$ km/s, the authors 
found $s_0=8\pm1\;h^{-1}$Mpc and $\gamma=1.9\pm0.7$.  
These numbers are very close to those given above for 
the P groups with $n\geq 3$ members. 

The fact that richer ($n\geq 5$ members) groups have a CF of greater amplitudes
than poorer groups ($n\geq 3$ members) is likely to be mostly due to 
an appreciable component of spurious groups among the latter, since spurious 
groups should cluster like field galaxies. As a matter of fact, N-body simulations 
showed that an appreciable fraction of the poorer groups, those with $n<5$ members, 
is false (i.e. unbound density fluctuations), whereas the richer groups almost always 
correspond to real systems (e.g., Ramella et al. 1997; Frederic 1995 a, b), as 
indicated also by optical and X-ray studies (e.g., Ramella et al. 1995; Mahdavi et 
al. 1997). Part of the difference in the degree of clustering of richer and poorer 
groups might be due to the existence of relationship between the strength of 
clustering and the system's richness. This relationship has been discussed for a 
variety of systems by several authors (e.g., Bahcall \& West 1992; Croft et al. 1997).  
The system's richness is usually taken as 
a rough tracer of the system's mass, which is the physical quantity related to 
the predictions of cosmological models. 

As regards the volume-limited samples of groups, power-law fits for the CF 
over the $3.5\leq s \leq 20\;h^{-1}$ Mpc range give, for instance, $s_0=
5.5\pm0.8\;h^{-1}$ Mpc, $\gamma=1.6\pm0.4$ ($s_0=5.6\pm0.8\;h^{-1}$ Mpc,
$\gamma=1.5\pm0.3$) for the H (P) groups with $n\geq 3$ members which are
characterized by a number density $n_g=8\cdot 10^{-4}\;h^3$ Mpc$^{-3}$ and a mean
intergroup separation $d= n_g^{-1/3}=11\;h^{-1}$ Mpc. Moreover, we obtain
$s_0=6.0\pm1.8\;h^{-1}$ Mpc, $\gamma=1.6\pm0.7$ ($s_0=7.3\pm1.4\;h^{-1}$ Mpc,
$\gamma=1.6\pm0.5$), on scales $3.5 - 30\;h^{-1}$ Mpc, for the H (P) groups with
$n\geq 5$ members, which are characterized by a number density $n_g=3\cdot
10^{-5}\;h^3$ Mpc$^{-3}$ and a mean intergroup separation $d=15\;h^{-1}$Mpc.  These
results confirm that $s_0$ tends to increase with the number of group members,
although there is only a marginal difference with respect the CF amplitude 
of the corresponding volume-limited sample of galaxies, for which a
power-law fit (over the $2.7 \leq s \leq 12\;h^{-1}$ Mpc range) gives
$s_0=5.09\pm0.10\;h^{-1}$ Mpc, $\gamma = 1.64\pm0.08$. 

The above-mentioned values of $s_0$ and $d$ for the volume-limited samples of groups 
are not far from the $s_0=0.4 d$ relationship proposed by Bahcall \& West (1992). 
Specifically, they  agree with the $s_0$ - $d$ curves predicted by N-body 
cosmological simulations which reproduce the observed $s_0$-$d$ relation 
of galaxy clusters (especially in the case of  low-density (open) CDM; cf.
Figure 8 of Governato et al. 1999).

\section{The Dependence of the Group Correlation Function on Group Properties}

In this section we wish to check on a possible dependence of the strength of 
clustering of groups on some properties of groups, such as the velocity dispersion, 
the virial radius, the mean pairwise separation of group members, the virial 
mass, the crossing time, and the morphological composition of group members.

\subsection{Velocity Dispersion}
First, we have computed the line-of-sight velocity dispersion $\sigma_v$,  
which for a spherical system is smaller by $\sqrt{3}$ than the 3D 
velocity dispersion. Specifically, we have estimated the "robust" velocity 
dispersion by using the biweight estimator for rich groups ($n\geq15$) and the 
gapper estimator for poorer groups (see the ROSTAT routines by Beers, Flynn,  
\& Gebhardt 1990). Beers et al. (1990) have discussed the superiority of these techniques 
in terms of efficiency and stability when systems with a small number of members 
are analyzed (cf. also Girardi et al. 1993). We have applied the relativistic 
correction and the standard correction for velocity errors (Danese, De Zotti, \& 
di Tullio 1980), adopting a typical velocity error of 30 km/s for each galaxy. 
This error is the average of the mean errors estimated in the RC3 catalog 
(de Vaucouleurs et al. 1991) for optical and radio radial velocities. For a few 
groups, corrections for velocity errors lead to $\sigma_v=0$.

In general, the $\sigma_v$-distribution of P groups is shifted to greater values
than that of H groups. The medians of $\sigma_v$ are 129 and 83 km/s for the
respective groups with $n\geq$3 members and tend to increase as we go to richer
groups (e.g. the respective medians are 161 and 106 km/s for the groups with 
$n\geq$ 5 members). The P and H LEDA groups identified by Garcia (1993) exhibit
a similar difference, which is due to differences in the algorithms of group selection 
(the $\sigma_v$-medians are 151 and 118 km/s for the P and H LEDA groups with
$|b|>20^{\circ}$; see Girardi \& Giuricin 2000). The $\sigma_v$-median of the NOG H
groups is in substantial agreement with the corresponding medians of other catalogs
of groups selected with the H algorithm, the PGC groups (73 km/s; see Gourgoulhon, 
Chamaraux, \& Fouqu\'e 1992) and NBG groups (100 km/s; see Tully 1987). 
As for the P algorithm, the
$\sigma_v$-median of the NOG P groups is close to the value of the revised CfA1
groups (116 km/s; Nolthenius 1993), in which a restrictive velocity link parameter 
was used, and quite smaller (as expected) than the medians
of $\sim$170 - 200 km/s relative to samples of groups in which generous velocity link 
parameters were adopted (e.g. the CfA2 north groups
by Ramella, Pisani, \& Geller 1997, the PPS groups by Trasarti-Battistoni 1998, the
ESP groups by Ramella et al. 1999, the SSRS2 groups by Ramella et al. 2000).

We divide the NOG P and H groups into two subsamples of equal sizes, i.e. groups
with $\sigma_v$ greater and smaller than the median values. In general, the redshift
distributions of the earlier group subsets are shifted to greater values than those
of the latter subsets, except in the case of P groups with $n\geq$3 members.
Therefore, for this group sample, we can compare the weighted CFs of the group subsets
relative to high and low $\sigma_v$- values (see Figure 13). High-$\sigma_v$ groups are
more clustered than their low $\sigma_v$ counterparts. Adopting the
above-mentioned rebinning procedure, we find $\xi(s^*) = 0.76^{+0.26}_{-0.28}$ and 
$1.61^{+0.41}_{-0.52}$ for the low-$\sigma_v$ and high-$\sigma_v$ groups, respectively.
This leads to a $\grsim 1\sigma$ significant difference.   
Using a power-law fit over the usual interval of separations we obtain $s_0= 8.6\pm0.8\;
h^{-1}$ Mpc, $\gamma=2.1\pm0.3$ for the high-$\sigma_v$ groups and $s_0=5.3\pm1.5\;h^{-1}$
Mpc, $\gamma= 1.4\pm0.5$ for the low-$\sigma_v$ groups.

Then we have subdivided the volume-limited samples of groups on the basis of 
the medians of $\sigma_v$. Table 6 reports the values of the $\xi$
evaluated at $s^*$ for the volume-limited H and P groups with $n\geq3$ and $n\geq4$.
In this table we denote by $\sigma_v^{\dag}$ the limiting value adopted for subdividing the 
each sample of groups into two subsamples, i.e. the groups with high and low $\sigma_v$. 
We confirm an excess of clustering (at the $\sim 1 - 2 
\sigma$ confidence level) for systems of relatively
high $\sigma_v$-values in the case of H groups. The volume-limited P groups (with
$n\geq3$ and $n\geq4$) show a weak tendency in the same sense, which is, however, not
statistically significant.

For these groups we have also calculated the distributions of pairwise group separations. 
We find that the two distributions relative to the high-$\sigma_v$ and low-$\sigma_v$ P groups with 
$n\geq 4$ members differ at the 99.9\% confidence level according to the 
KS test; in other words,   
high-$\sigma_v$ groups have a distribution shifted to lower values than the other  
groups, as is expected from their greater degree of clustering. For the H groups 
the corresponding difference is significant at the 98.3\% confidence level. 
No appreciable difference (i.e. no difference at the $>90$\% confidence level) is seen 
for the P and H groups with $n\geq 3$ members.

\subsection{Virial Radius and Mean Pairwise Separation}
We have computed the projected virial radius $R_{pv}$ as 
$R_{pv} = n(n-1)/(\sum_{i>j} R_{ij}^{-1})$ where  
$R_{ij}$ is the projected distance between galaxies and $n$ the number of group 
members. For a spherical system the projected virial radius is smaller by $\pi/2$ 
than the virial radius $R_v$. We have computed the mean pairwise member separation 
$R_m$ as $(4/\pi) <R_{ij}>$ where $4/\pi$ is the projection factor. $R_m$ is likely 
to be a more robust estimate of the group size than the virial radius.  

The H groups have typically greater values of $R_m$ and $R_{pv}$ than the P groups. 
The $R_{pv}$($R_m$)-medians are 0.60 (0.65) and 0.48 (0.50) $\;h^{-1}$ Mpc for the
respective groups with $n\geq$3 members and tend to increase as we go to richer
groups (e.g., the respective medians are 0.66 (0.78) and 0.52 (0.57) $h^{-1}$ 
Mpc for the groups with $n\geq 5$ members).
The NOG P groups have smaller sizes
than the LEDA P groups (the LEDA H and P groups have a $R_{pv}$-median of 
0.62 $h^{-1}$ Mpc; see Girardi \& Giuricin 2000) and the new UZC groups 
by Pisani et al. 2001 (which have a $R_{pv}$-median of $0.64 h^{-1}$ Mpc),  
probably because we have adopted smaller link parameters. Other catalogs of 
H and P groups are not easily comparable to the NOG groups in this respect, because 
in general the authors (e.g. Ramella et al. 1997) calculate the harmonic 
radius, which is systematically smaller than the virial radius by a factor of 
$\sim$ 2 (as discussed by Giuricin 1989). 

Since the redshift distributions of the high-$R_{pv}$ and high-$R_m$ groups is
considerably shifted to greater values than that of the low-$R_{pv}$ and low-$R_m$
groups, we consider only the volume-limited samples of groups in the analysis of the
dependence of CF on $R_{pv}$ and $R_m$. 

Subdividing the volume-limited samples of groups according to the medians of
$R_{pv}$ and $R_m$, we find that the high-$R_{pv}$ and high-$R_m$ groups (with
$n\geq3$, $n\geq4$, $n\geq$5) tend to be more clustered than the low-$R_{pv}$ and
low-$R_m$ counterparts. As in the case of $\sigma_v$, the effect is more pronounced
for H groups than for P groups. Moreover, the effect gets stronger if we increase
the limiting value of $R_{pv}$ ($R_m$) so as to subdivide the groups into a
high-$R_{pv}$ ($R_m$) sample containing $\sim$1/3 of groups and a low-$R_{pv}$
(low-$R_m$) sample comprising $\sim2/3$ of groups. Figure 14  shows the $\xi(s)$
of the high-$R_{pv}$ and low-$R_{pv}$ H groups (with $n\geq$4). Table 7
contains the values of $\xi(s^*)$ calculated after rebinning. In this table
$R_{pv}^{\dag}$ and $R_m^{\dag}$ are the limiting values adopted for subdividing
each sample of groups into two subsamples of high and low size.  Table 7 reveals
that the statistical significance of this effect is at the $1 - 2\sigma$ level. For
the volume-limited samples of P and H groups, this effect is respectively slightly
larger and smaller than the effect related to $\sigma_v$. 
 
\subsection{Virial Mass} 
We have calculated the virial mass $M_v$ as $M_v = 3\cdot
(\pi/2) (\sigma_v)^2 R_{pv}/G$. ($M_v$ has not been calculated 
in systems having $\sigma_v=0$). 
For a system which is virialized and which has a mass
distribution similar to the galaxy distribution, the virial mass is a reliable
indicator of the mass. 

Reflecting  the difference in $\sigma_v$, the P groups 
are, on average, more massive than the H groups. The $M_v$-medians are $8.7\cdot 
10^{12}\;h^{-1}\;M_{\odot}$ and $5\cdot 10^{12}\;h^{-1}\;M_{\odot}$ for the 
respective groups with $n\geq$3 and increase as we go to richer systems 
(e.g. the respective medians are $1.4\;h^{-1}\;10^{13}\;M_{\odot}$ and 
$7.8\;h^{-1}\;M_{\odot}$ for the gropus with $n\geq$ 5 members). 
The P and H LEDA groups exhibit an even larger difference in the same sense 
between the $M_v$-medians ($14.8\cdot 10^{12}\;h^{-1}\;M_{\odot}$ and 
$6.6\cdot 10^{12}\;h^{-1}\;M_{\odot}$, respectively; see Girardi \& Giuricin 
2000), because of differences in the typical sizes of P and H groups discussed 
in the previous subsection.  

As in the case of $R_{pv}$ and $R_m$, since the redshift
distribution is related to $M_v$, we consider only the volume-limited samples of
groups in examining the dependence of CF on $M_v$. 

Subdividing the groups according to the $M_v$-medians, we find that the more massive
groups tend to be more clustered than the less massive ones, that this effect is
more pronounced for the H groups than for the P groups and gets stronger if we raise
the limiting values of $M_v$ above their medians. These results are parallel to
those found for $\sigma_v$ and (especially) for $R_{pv}$, as expected from the
dependence of $M_v$ on these two quantities. Figures 15 and 16 illustrates the
CFs of the high-$M_v$ and low-${M_v}$ groups. Table 8, which reports the values
of $\xi(s^*)$ for P and H groups with high and low $M_v$-values (compared to the
value denoted by $M_v^{\dag}$) and $n\geq$ 3 and 4 members, reveals that this effect
is in general not stronger than the effect on $\sigma_v$ and on $R_{pv}$ for the
volume-limited H and P groups, respectively. 

The comparison of the distributions of  pairwise group separations 
of the above-mentioned samples of groups reveals that high-$M_v$ groups 
have always lower separations, on average, than the corresponding 
low-$M_v$ groups, with a statistical significance ranging from 95.1\% 
to 99.99\% confidence level.

Most of galaxy groups are indeed in the phase of collapse and not yet virialized
(e.g. Giuricin et al. 1988, Pisani et al. 1992). Corrections for non-virialization
effects lead to masses which are greater than virial masses by $\sim$20-60\%,
depending on the group catalog and on the assumed cosmology (Girardi \& Giuricin
2000). However, the corrected masses correlate very well with the (uncorrected)
virial masses.  Therefore, these corrections has no effect on the analysis described
in this subsection. 

\subsection{Virial Crossing Time} 

We have calculated the virial crossing time $t_{cr}$ (expressed in units of ther
Hubble time $t_H$) as $t_{cr} = (3/5)^{3/2} [\pi R_{pv}/2]/[\sqrt{3} \sigma_v]$ (see
Giuricin et al. 1988 for further details). Owing to the difference in the values of
$\sigma_v$ and $R_{pv}$, the H groups have, in general, larger $t_{cr}$-values than
the P groups. The $t_{cr}$-medians are 0.29 and 0.18 for all H and P groups,
respectively, with the redshift distribution of the high-$t_{cr}$ P groups being
shifted to greater values with respect to that of low-$t_{cr}$ groups.  We have
detected no appreciable difference between the $\xi(s)$ of high- and low-$t_{cr}$ H
and P groups, for the total and volume-limited samples. Clearly, since 
$t_{cr}\propto R_{pv}/\sigma_v$, the above-mentioned effects on $R_{pv}$ and 
$\sigma_v$ cancel out in the ratio.  

\subsection{Morphological Composition}
We have calculated the fraction of early-type (E+S0) galaxies for the NOG groups. 
The median fraction is 0.2 for the total and volume-limited samples of H and 
P groups. As expected from the morphological segregation, this fraction is greater 
than the fraction of 15\% which refers to the whole NOG. 

Groups having a small fraction of early-type galaxies are in general 
nearer than the norm. We have detected no appreciable difference between the 
CFs of the H and P groups with small and large fractions of early-type 
galaxies for the total and volume-limited samples.

\section{Summary and Conclusions}

Using the two-point correlation function in redshift space, we study the 
clustering of the galaxies and groups of the Nearby Optical Galaxy (NOG) 
sample, which is a complete, distance-limited ($cz\leq 6000$ km/s) and 
magnitude-limited sample of $\sim$7000 nearby and bright optical galaxies, 
which covers 2/3 of the sky ($|b|>20^{\circ}$ ). We consider two catalogs 
of NOG groups identified through the hierarchical and percolation 
algorithms.  

Our main results can be summarized as follows:

1) The redshift-space two-point correlation function of NOG galaxies can be well 
described by a power-law, with slope $\gamma= 1.5$ and correlation length 
$s_0=6.4\;h^{-1}$ Mpc between 2.7 and $12\;h^{-1}$ Mpc, in substantial 
agreement with the results of most redshift 
surveys of optical galaxies. Optical surveys, characterized by 
quite different geometries, volumes, and selection criteria, 
exhibit similar galaxy correlation functions 
(at least between 2 and $20\;h^{-1}$Mpc), 
although they show significant discrepancies in the shape and 
normalization of the resulting 
galaxy luminosity functions (as discussed in Paper II).
The agreement between galaxy correlation functions derived 
for a wide range of volumes and sample radii (as defined in Guzzo 1997)  
is in contrast with the fractal interpretation of the galaxy 
distribution in the universe.

2) Subdividing the NOG into early-type (E-S0)  and late-type (spirals and 
irregular) galaxies, we note a pronounced morphological segregation, out to 
20$h^{-1}$ Mpc,  between the former objects (characterized by  
$s_0\sim 11\;h^{-1}$ Mpc and $\gamma\sim1.5$) and  the latter ones 
(characterized by $s_0\sim 5.6\;h^{-1}$ Mpc and $\gamma\sim1.5$), 
in qualitative agreement with many previous investigations. Remarkably, 
the morphological segregation persists also at relatively low luminosities 
($M_B> -19$ mag).

Subdividing further the NOG into several  morphological 
subtypes, we note a gradual increase of the strength of clustering from 
the late-type spirals to the S0s, especially on intermediate scales 
(around 5$h^{-1}$ Mpc).  

The relative bias factor ($\sim$1.7 in redshift space)  
between early- and late-type objects, for which there is no 
good agreement in the literature,  appears to   
be substantially constant with scale. However, this result is not incompatible  
with a possible scale dependence in the relation between galaxy 
number density and mass density (see, e.g., Blanton et al. 1999).

3) Analyzing different volume-limited samples, we find that the 
luminous galaxies turn out to be more clustered than the dim objects.
The luminosity segregation, which is significant for both early- and late-type 
objects, starts to become appreciable only at relatively high luminosities 
($M_B\lesssim -19.5$, i.e. $L\grsim 0.6\;L^{*}$) and is 
independent on scale (at least up to 10$h^{-1}$ Mpc). 

The very luminous galaxies ($M_B\leq$; $L\grsim 2.4\;L^{*}$) 
reside preferentially in binaries and groups (though not in clusters)  
and their degree of clustering ($s_0\sim 12\;h^{-1}$ Mpc) is  compatible  
with that of rich groups. 

Both morphological and luminosity segregations are two separate effects 
(one effect is not generated by the other). 
The fact that they are present also on large scales favors the interpretation 
that, on scales greater than $\sim 1\;h^{-1}$ Mpc,  the bulk of these effects is likely 
to be mostly primordial in origin, i.e. inherent in schemes of 
biased galaxy formation (e.g. Bardeen et al. 1986) and not induced by late environmental 
effects (e.g. due to merging, harassment, tidal stripping, accretion, ram pressure).

4) The NOG groups identified with the H and P algorithms, though having somewhat
different distributions of velocity dispersions, virial masses, and sizes, have
similar clustering properties. The groups show a degree of clustering which is
intermediate between those of galaxies and clusters. Compared to galaxies, they
exhibit an excess of clustering by a factor $\sim$1.5 and $\sim$2, for groups with
$n\geq3$ and $\geq5$ members, respectively. The groups with $n\geq3$ members are
characterized by $s_0\sim 8\;h^{-1}$ Mpc, which should be regarded as a lower limit
since poor groups can be appreciably contaminated by spurious groups. The groups
with $n\geq5$ members, which have typical (median) velocity dispersions of
$\sigma_v\sim 100 - 150$ km/s, virial radii of $R_v\sim 1\;h^{-1}$ Mpc , and virial
masses of $M_v\sim 10^{13}\;h^{-1}\;M_{\odot}$, are characterized by $s_0\sim
10\;h^{-1}$ Mpc. 

5) The strength of group clustering depends on the physical properties of 
groups. Specifically, groups with greater velocity dispersions, sizes 
(as traced by virial radii and mean pairwise member separations), and 
virial masses tend to be more clustered than those with lower values of 
these quantities. The H groups display a slightly stronger effect than the P 
groups. On the other hand, there is no difference in the 
degree of clustering between groups with small and large proportion 
of early-type galaxies.

In comparing the results from different samples through the analysis of 
correlation functions in redshift space, we have adopted the working 
hypothesis that, to first order, the influence of redshift space 
distortions will affect the samples being compared in similar ways.
We are undertaking the analysis of the real-space 
clustering in the NOG. The results of this work, which is in progress, will  
be presented in a forthcoming paper.

Our results can provide useful constraints to current models of galaxy formation,
which are able to fit the two-point galaxy correlation function in a $\lambda$CDM
cosmology (e.g., Benson et al.  2000) and also attempt to reproduce, with a limited
success, the observed morphological and luminosity segregation (e.g., Kauffmann et
al. 1999).  As for the clustering of groups, our findings can offer some additional
constrains for N-body simulations which in different cosmologies predict the
clustering properties of galaxy systems (see, e.g., Governato et al. 1999 and
Colberg et al. 2000 for recent relevant accounts).

\acknowledgments

We are grateful to L. Guzzo, who have sent us some data, and 
we thank S. Borgani, F. Mardirossian, P. Monaco, M. Ramella 
for interesting conversations. 

S. S. acknowledges receipt of a grant of the University of Trieste and a 
TRIL fellowship of the International Center for Theoretical Physics of Trieste.

This work has been partially supported by the Italian Ministry of University, 
Scientific and Technological Research (MURST) and by the Italian Space Agency 
(ASI).

\newpage

\figcaption[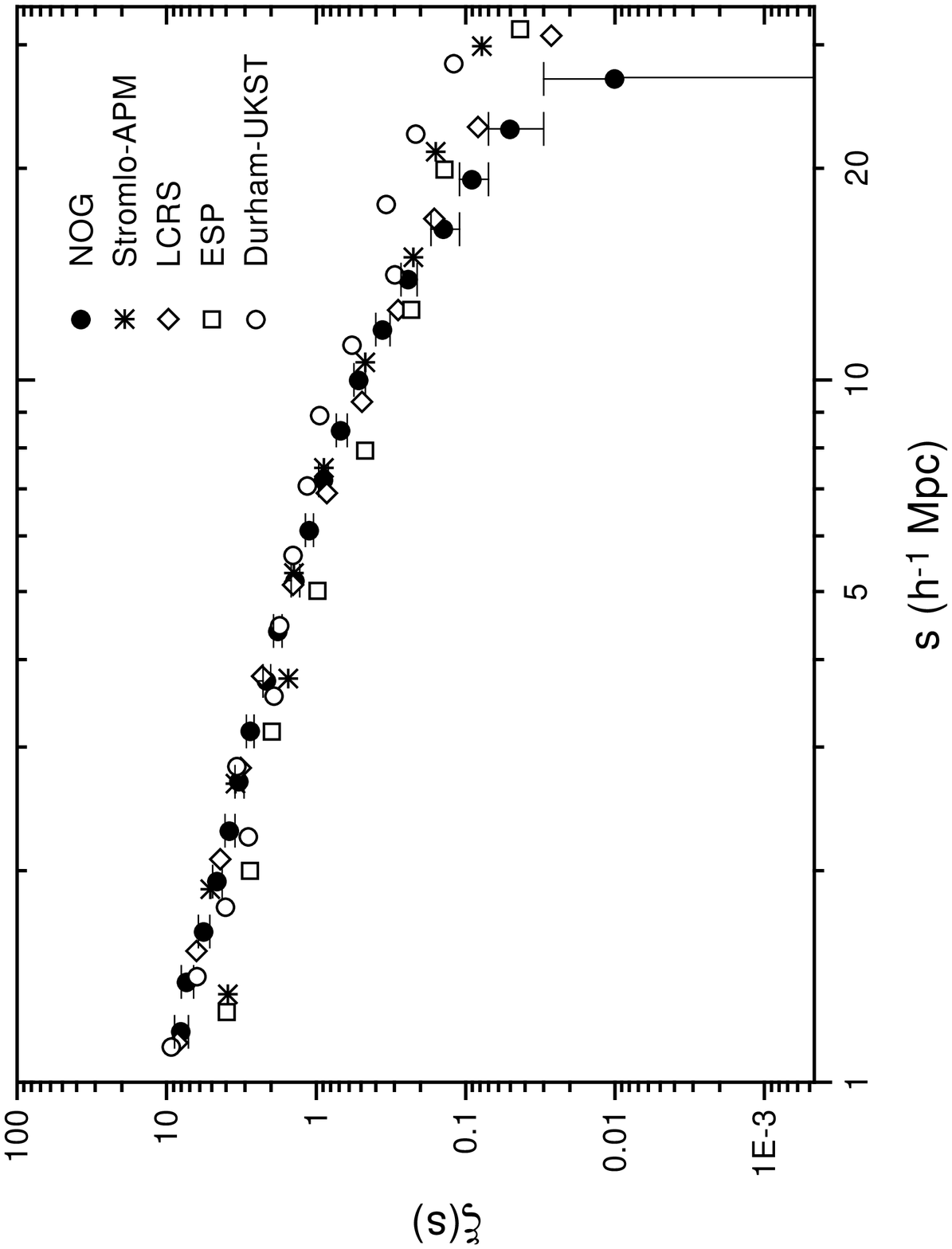]{The redshift-space galaxy correlation function from the whole NOG,  
compared to results from the Las Campanas (LCRS), Stromlo-APM, Durham-UKST, 
and ESP redshift surveys; 1 $\sigma$ error bars are shown for the 
NOG correlation function.}

\figcaption[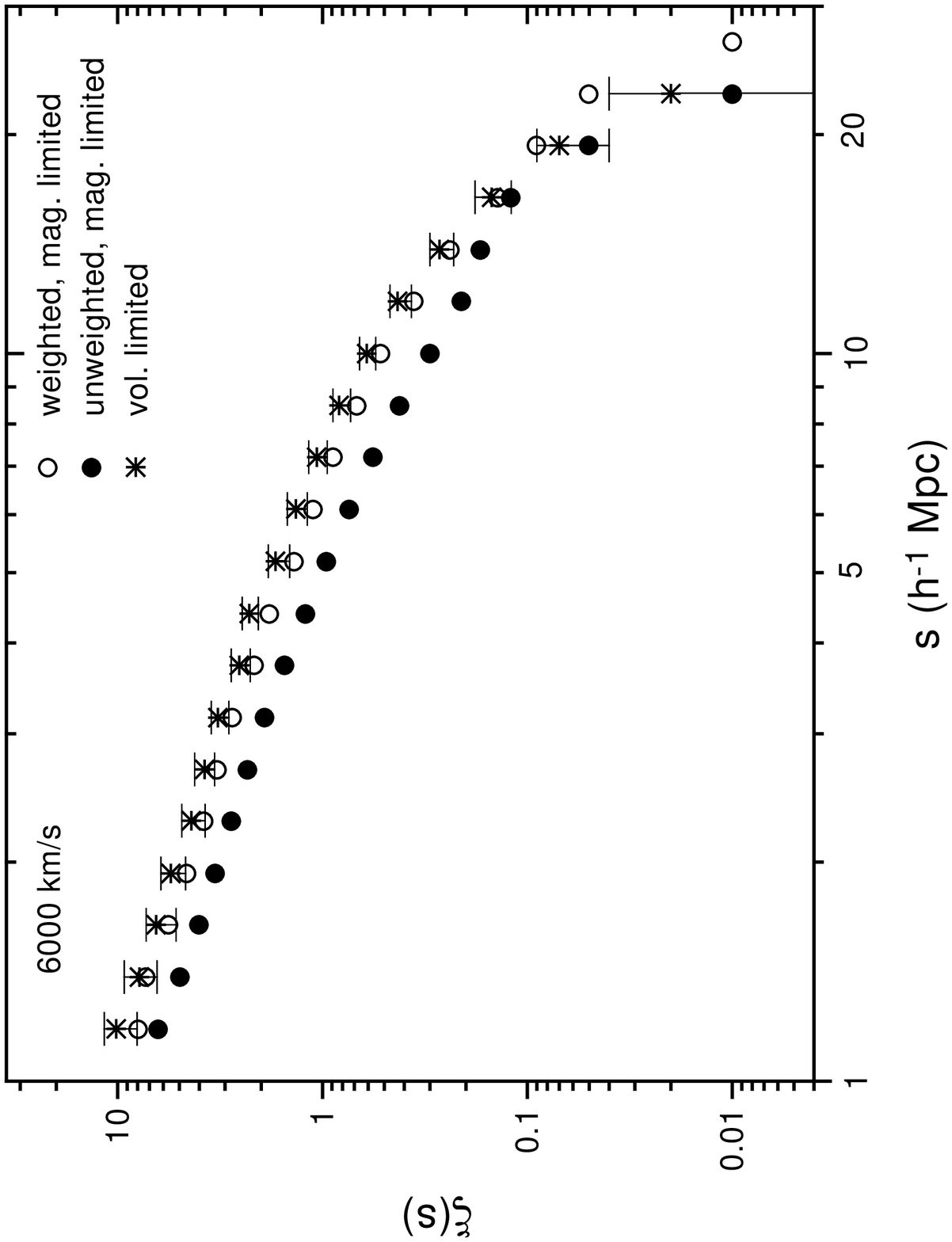]{Comparison between the weighted (open circles) and 
unweighted (dots) correlation functions for the whole NOG, and the 
correlation function for the volume-limited sample with depth of 6000 km/s 
(stars). For the sake of clarity, error bars are shown for the last case only.}

\figcaption[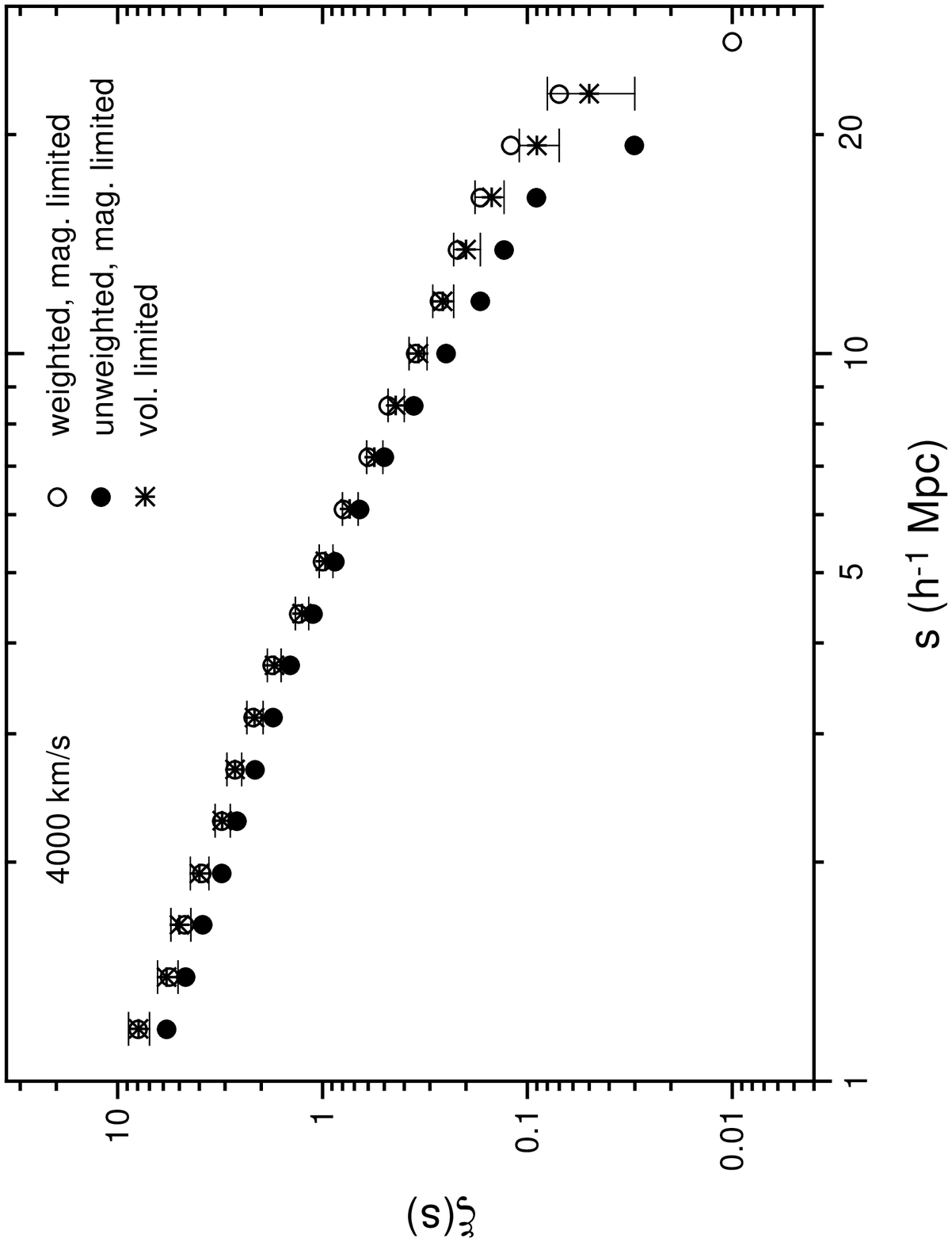]{The same as in Figure 3, but for samples limited at 
the depth of 4000 km/s.}

\figcaption[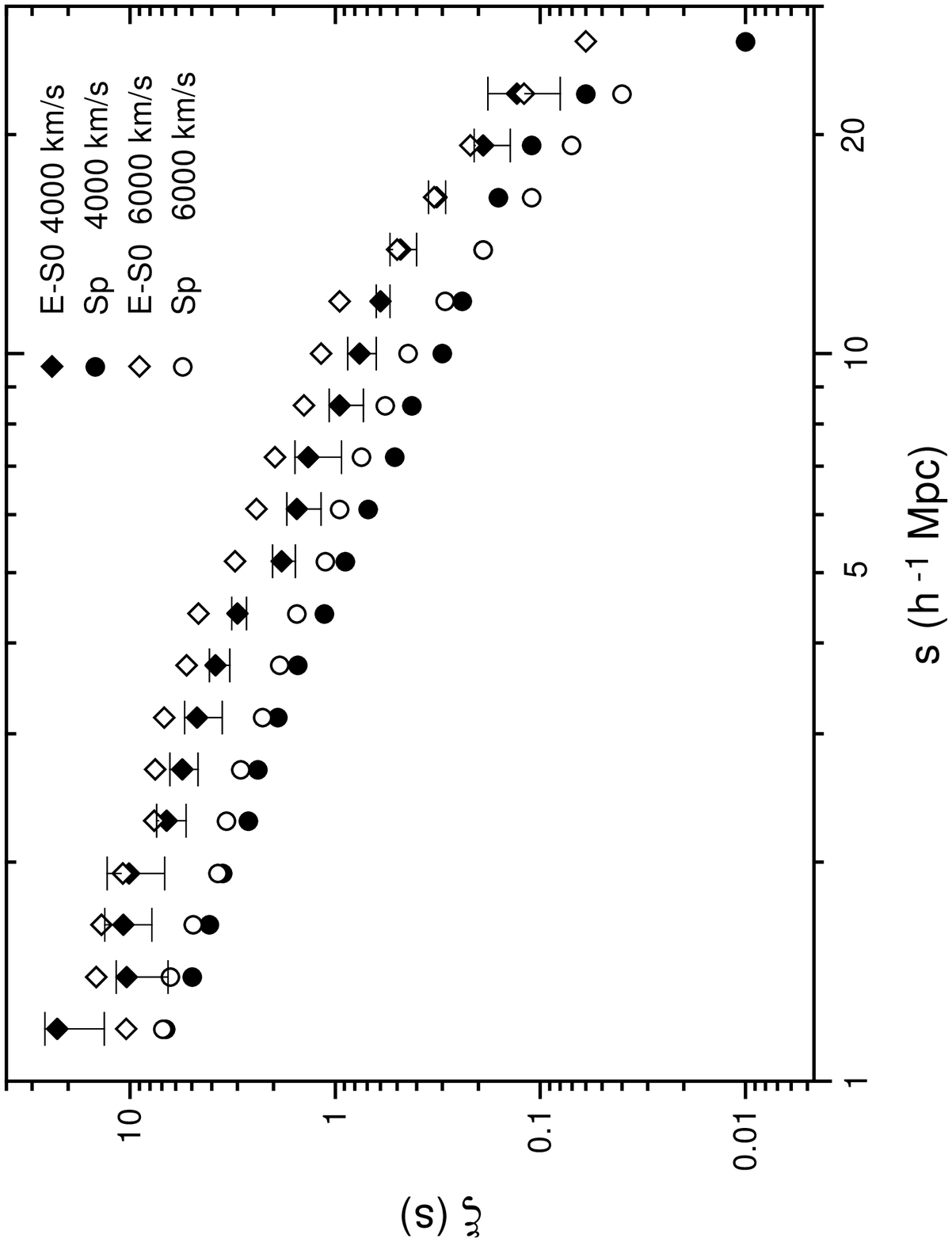]{Comparison of the correlation functions for the 
early-type (open diamonds) and late-type (open circles) galaxies of 
the whole NOG, and for the early-type (filled diamonds) and late-type 
(dots) galaxies of the magnitude-limited sample truncated at 4000 km/s.
For the sake of clarity, errors bars are shown for one sample only.}

\figcaption[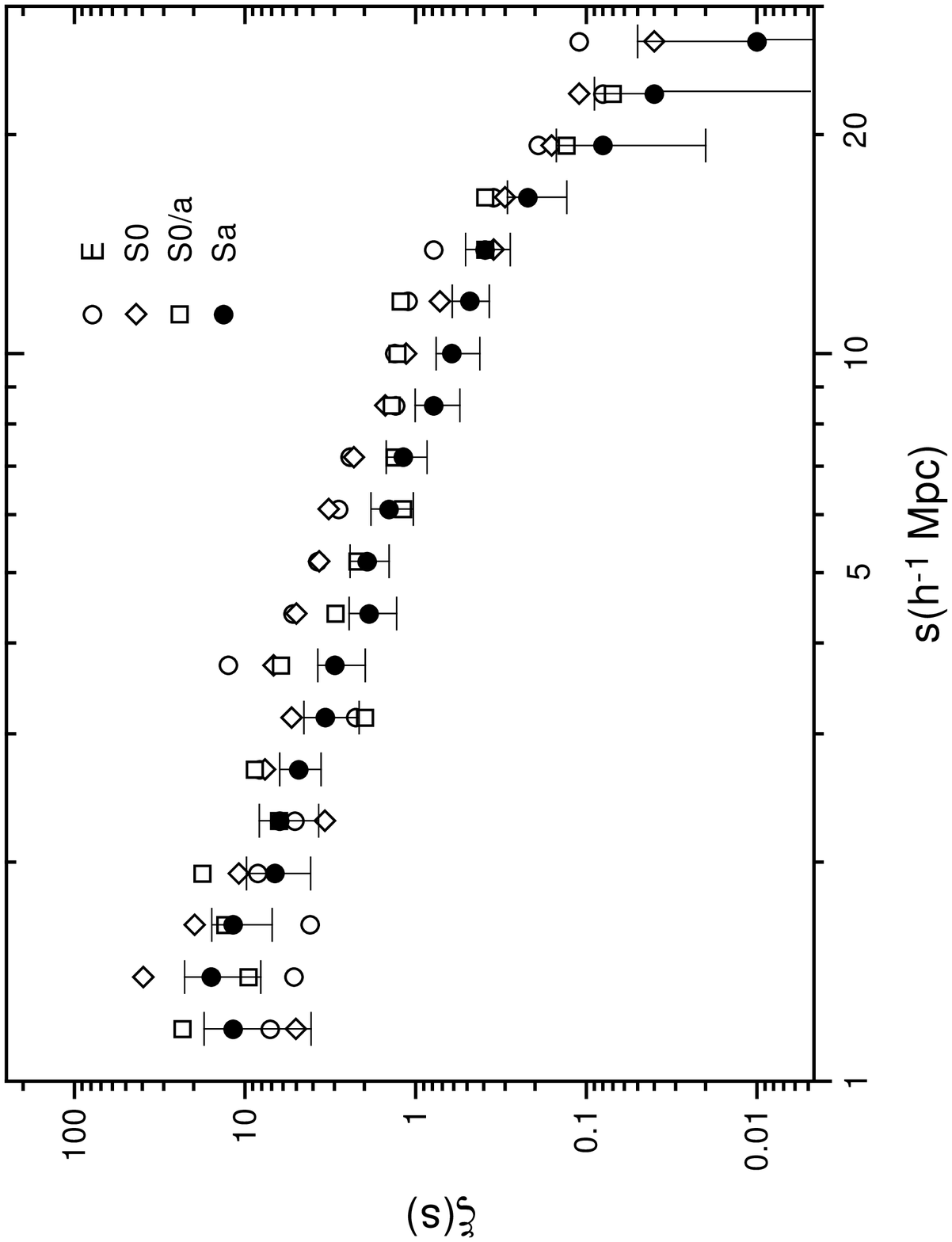]{Comparison of the correlation functions for  
the E (open circles), S0 (open diamonds), S0/a (open squares), 
and Sa (dots) morphological types of the whole NOG. 
For the sake of clarity, error bars are shown for the last 
sample only.}

\figcaption[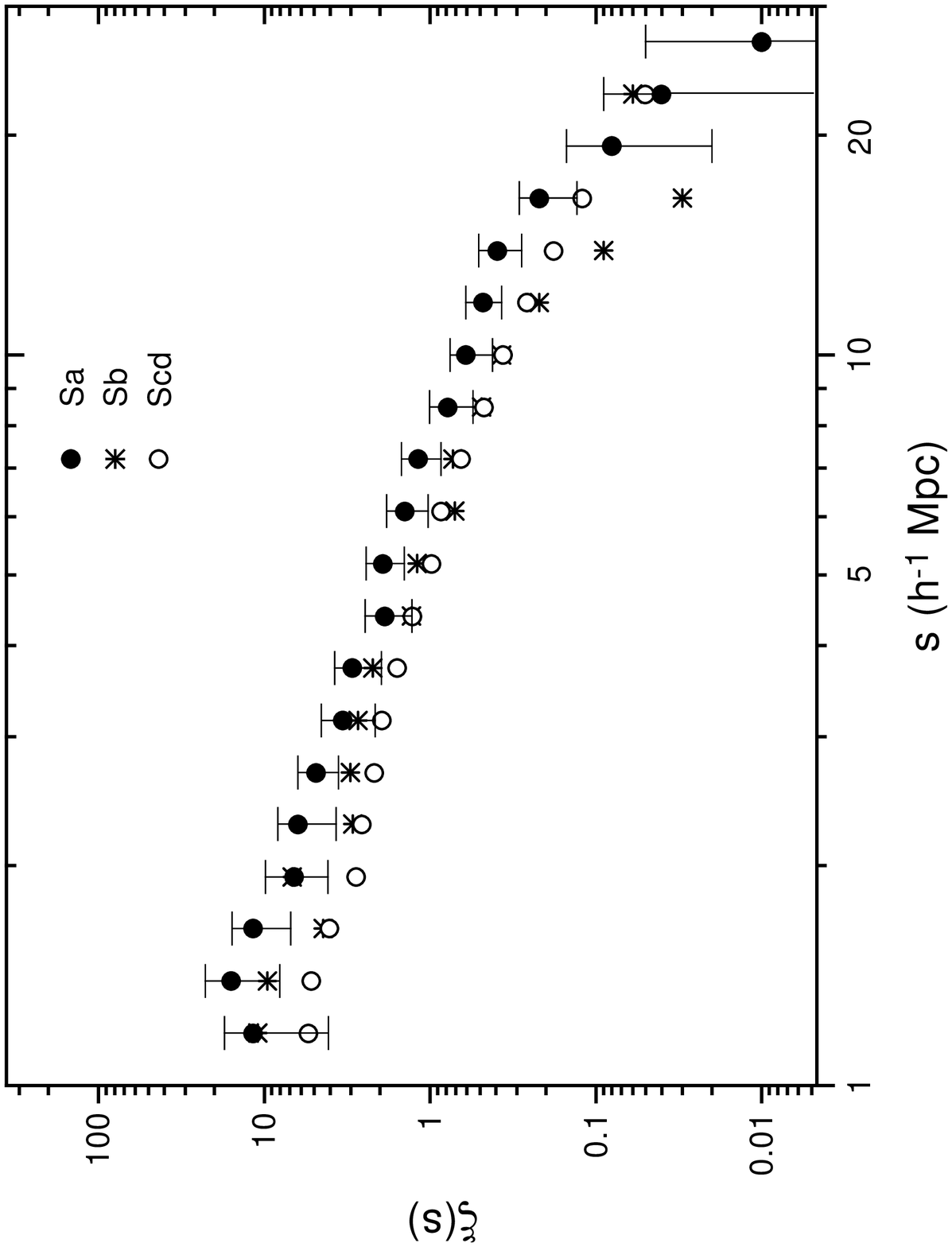]{Comparison of the correlation functions for 
the Sa (dots), Sb (stars), and Scd (open circles) 
morphological types of the whole NOG. For the sake of clarity, 
error bars are shown for the first sample only.}

\figcaption[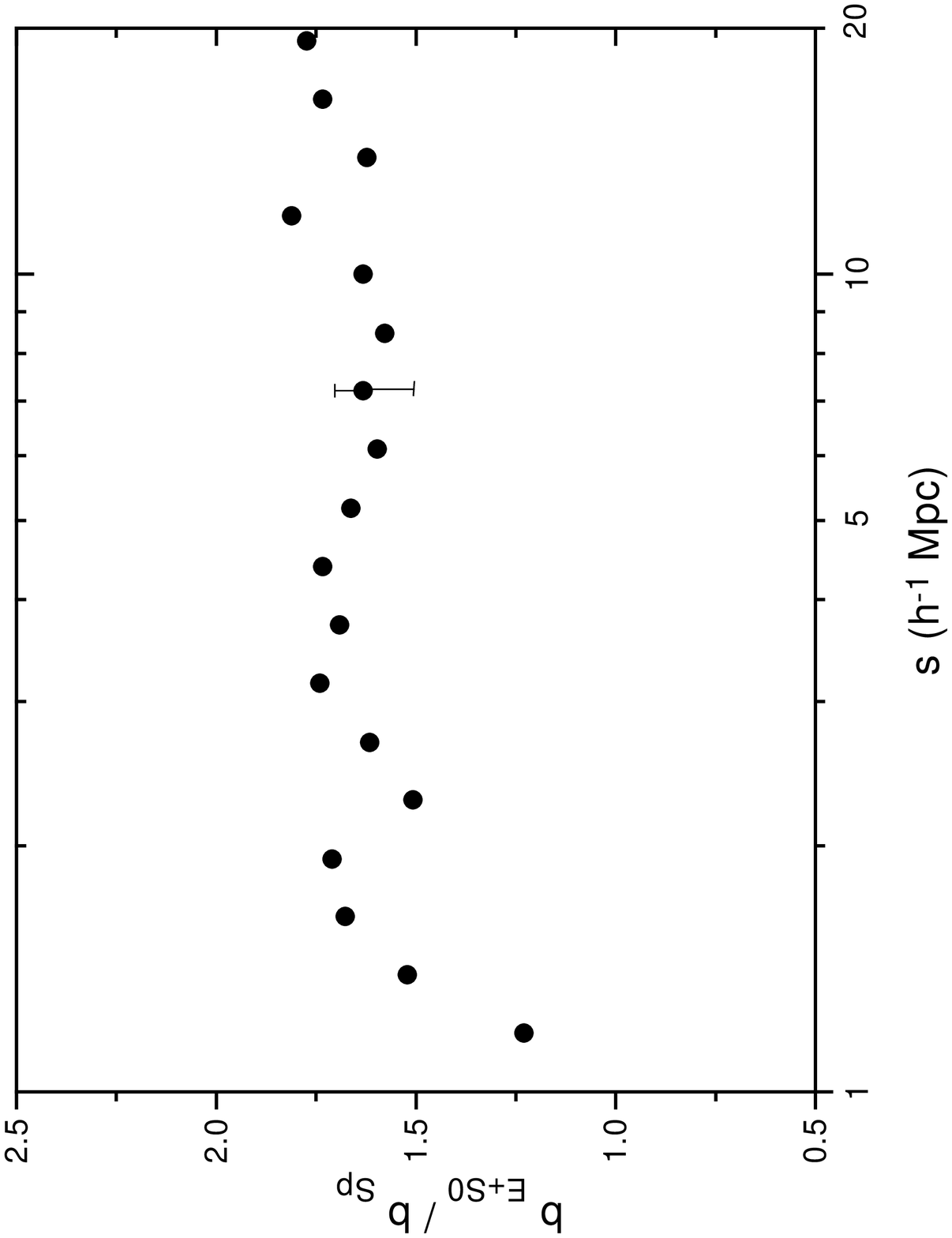]{The relative bias between early- and late-type 
galaxies in the NOG. A typical error bar is shown.}

\figcaption[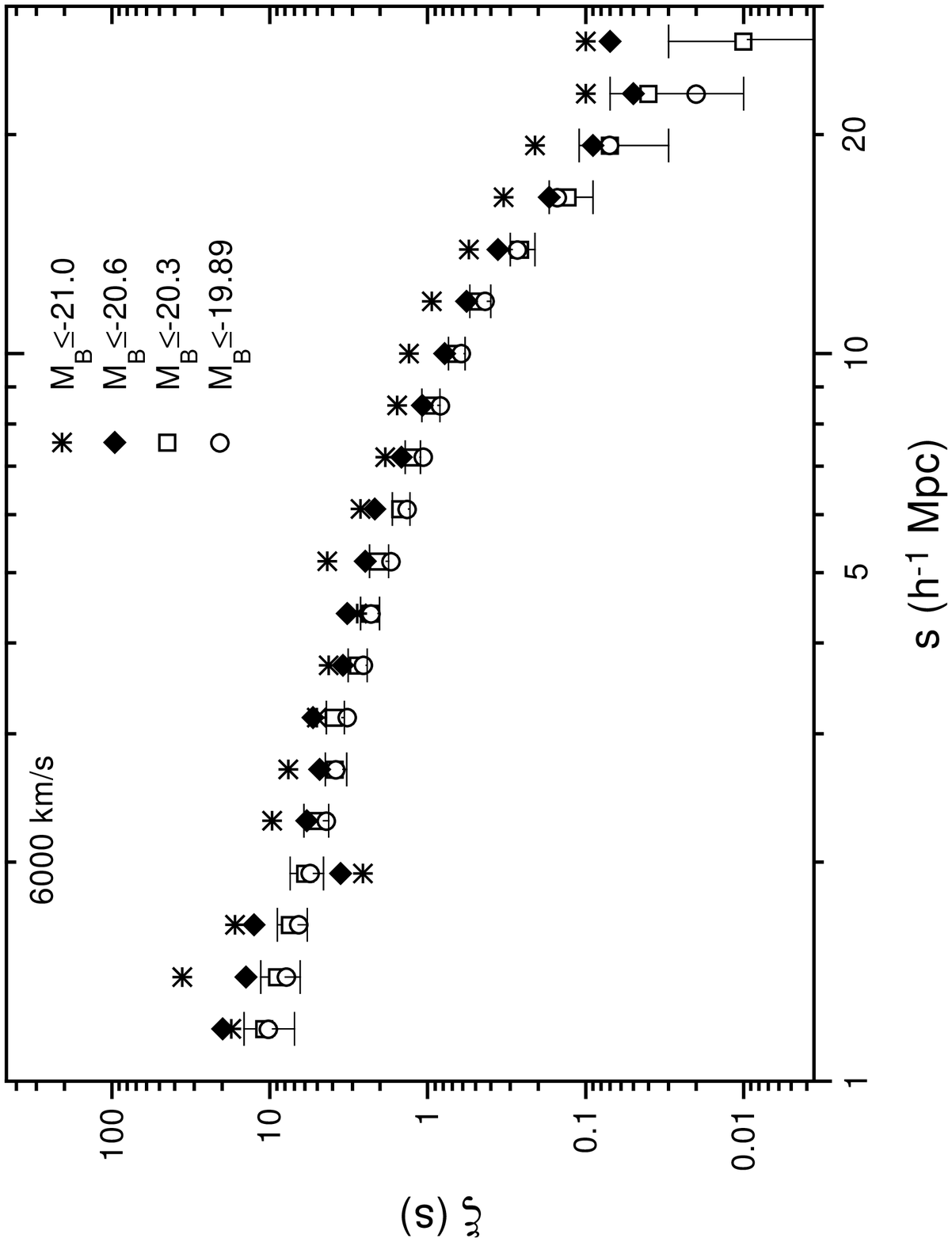]{Comparison of the correlation functions for 
different subsamples of the volume-limited sample with depth of 
6000 km/s. Each subsample corresponds to a given luminosity class.
Error bars are shown for one subsample only.}

\figcaption[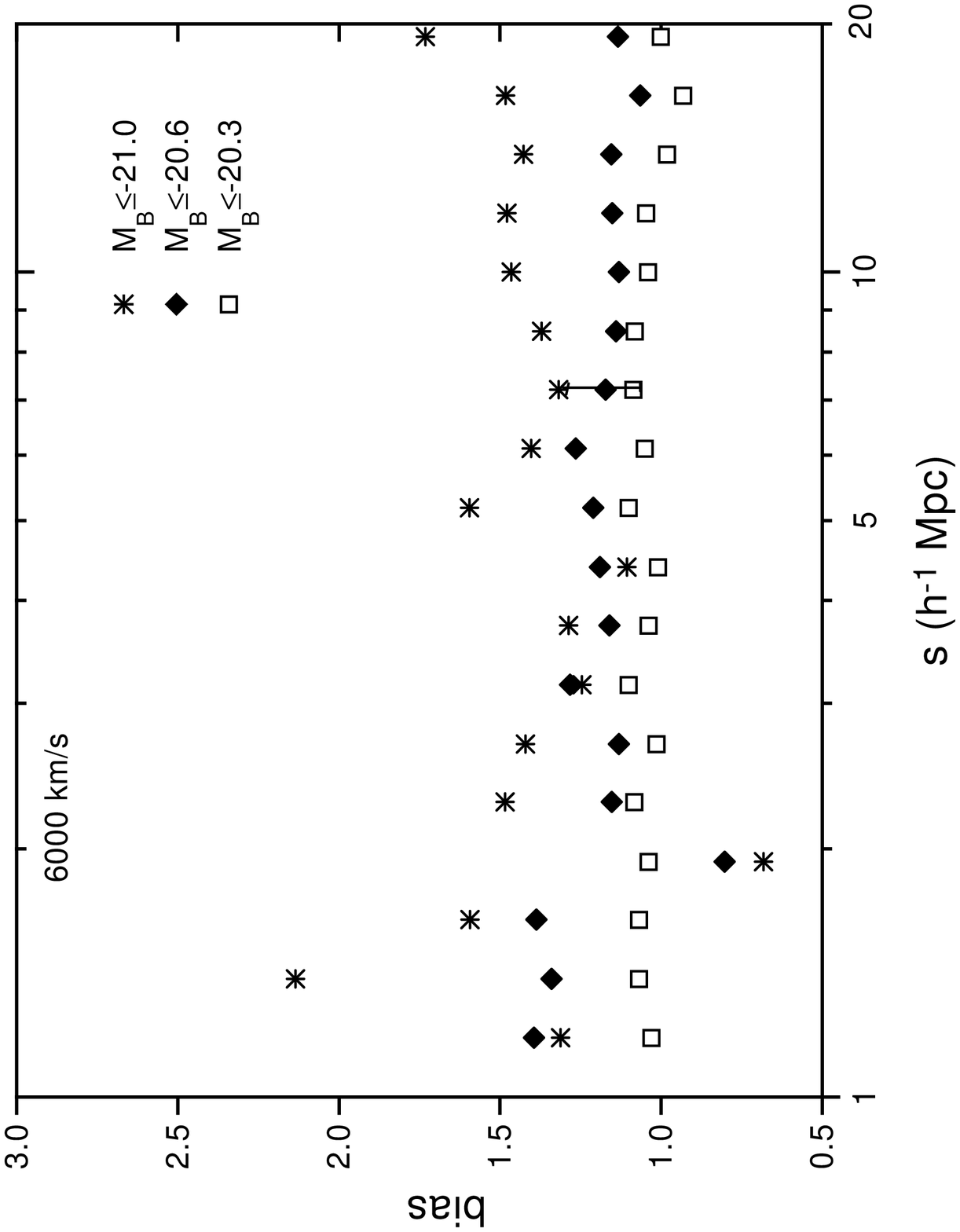]{The relative bias between the subsamples of galaxies 
of different luminosity classes considered  in 
Figure 8. The bias is normalized to the value 
relative to the faintest subsample ($M_B\leq -19.89$ mag). 
A typical error bar is shown.}

\figcaption[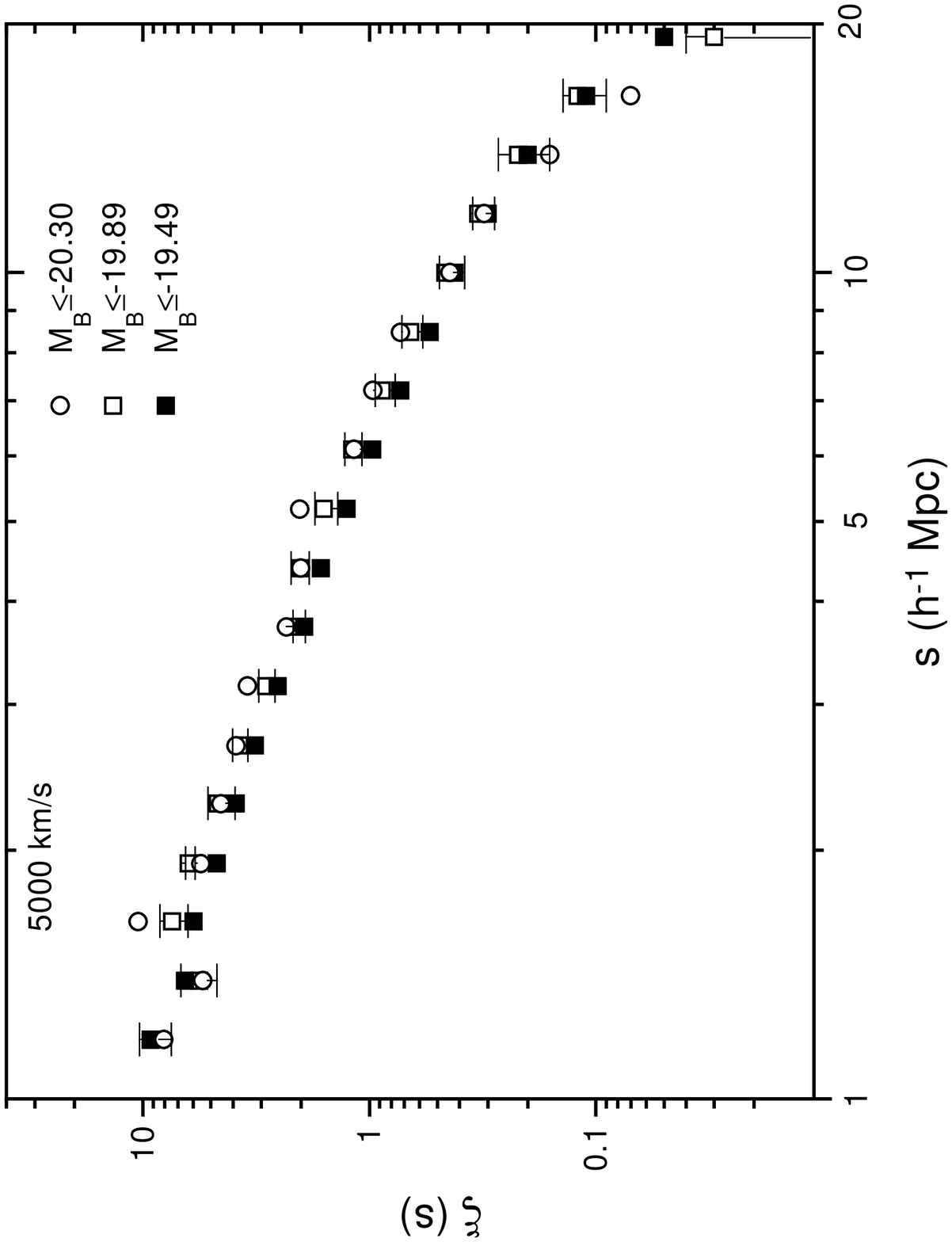]{Comparison of the correlation functions for 
different subsamples of the volume-limited sample with depth of 
5000 km/s. Each subsample corresponds to a given luminosity class.
Error bars are shown for one subsample only.}

\figcaption[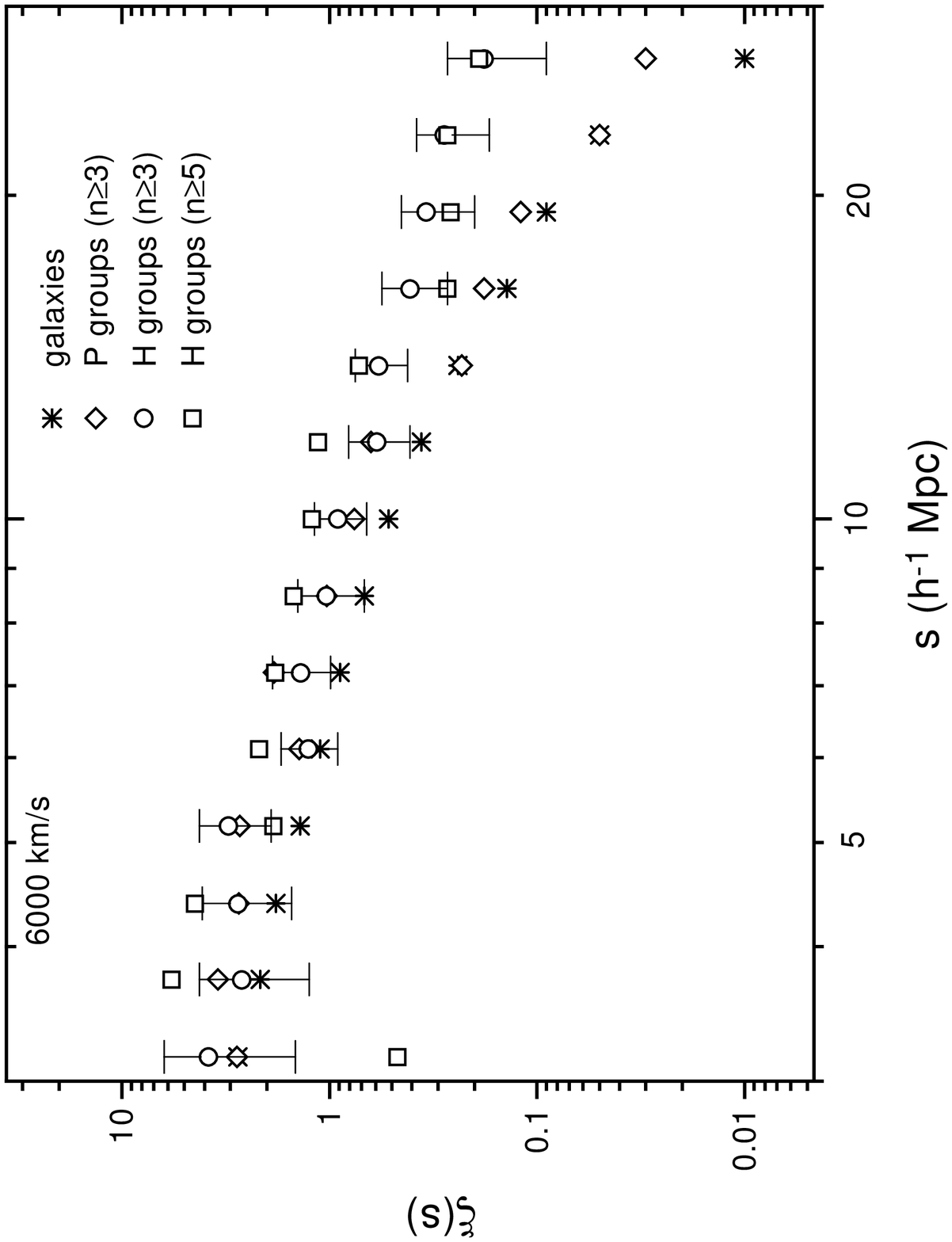]{Comparison of the correlation functions for 
the NOG galaxies (stars), the P groups with $n\geq 3$ members 
(diamonds), the H groups with $n\geq 3$ members (open circles), 
and the P groups with $n\geq 5$ members (open squares).
Error bars are shown for one case only.}

\figcaption[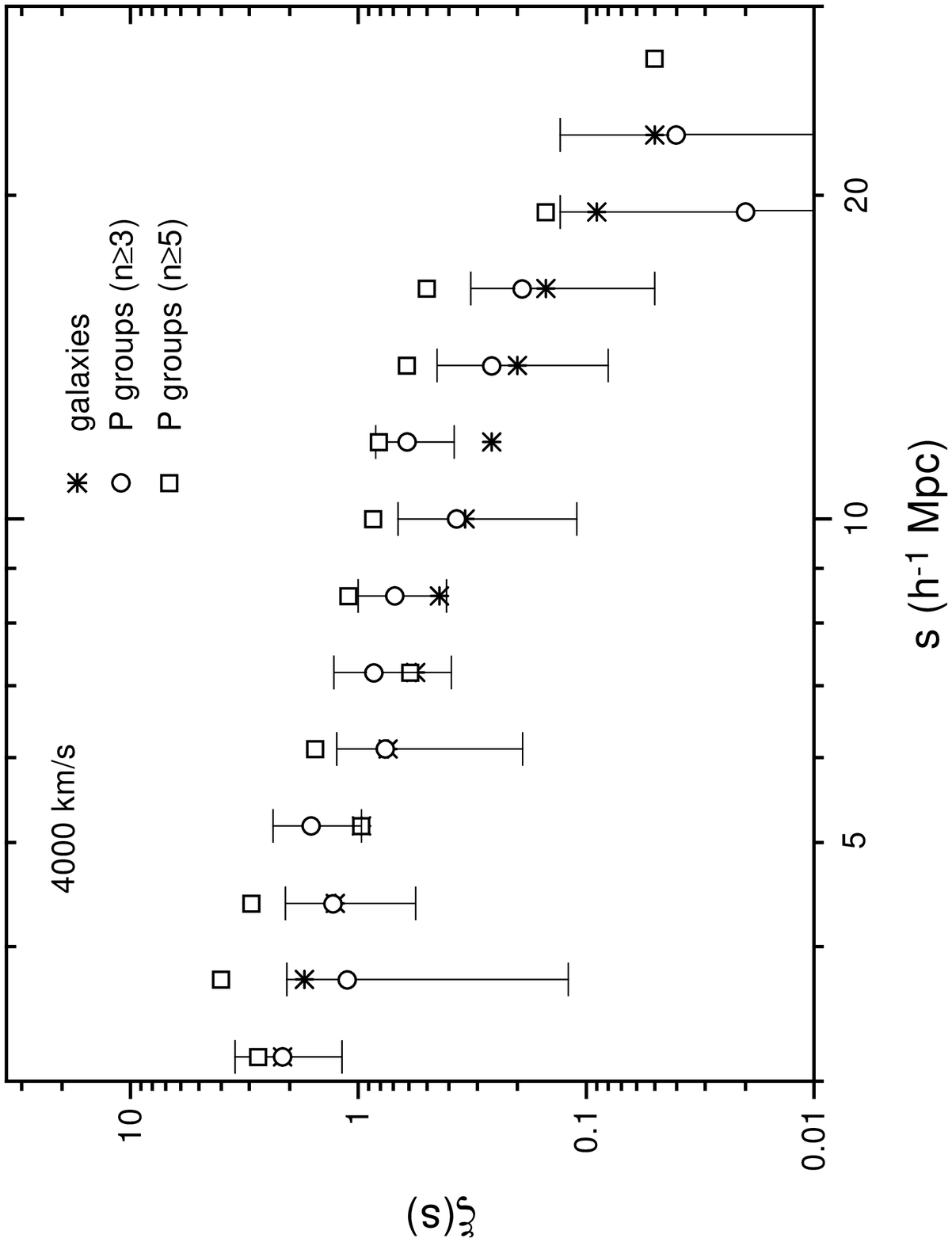]{Comparison between the correlation functions 
for the the volume-limited samples (with depth of 4000 km/s) 
of galaxies (stars) and P groups with $n\geq 3$ 
(open circles) and $n\geq 5$ members (open squares).}  
 
\figcaption[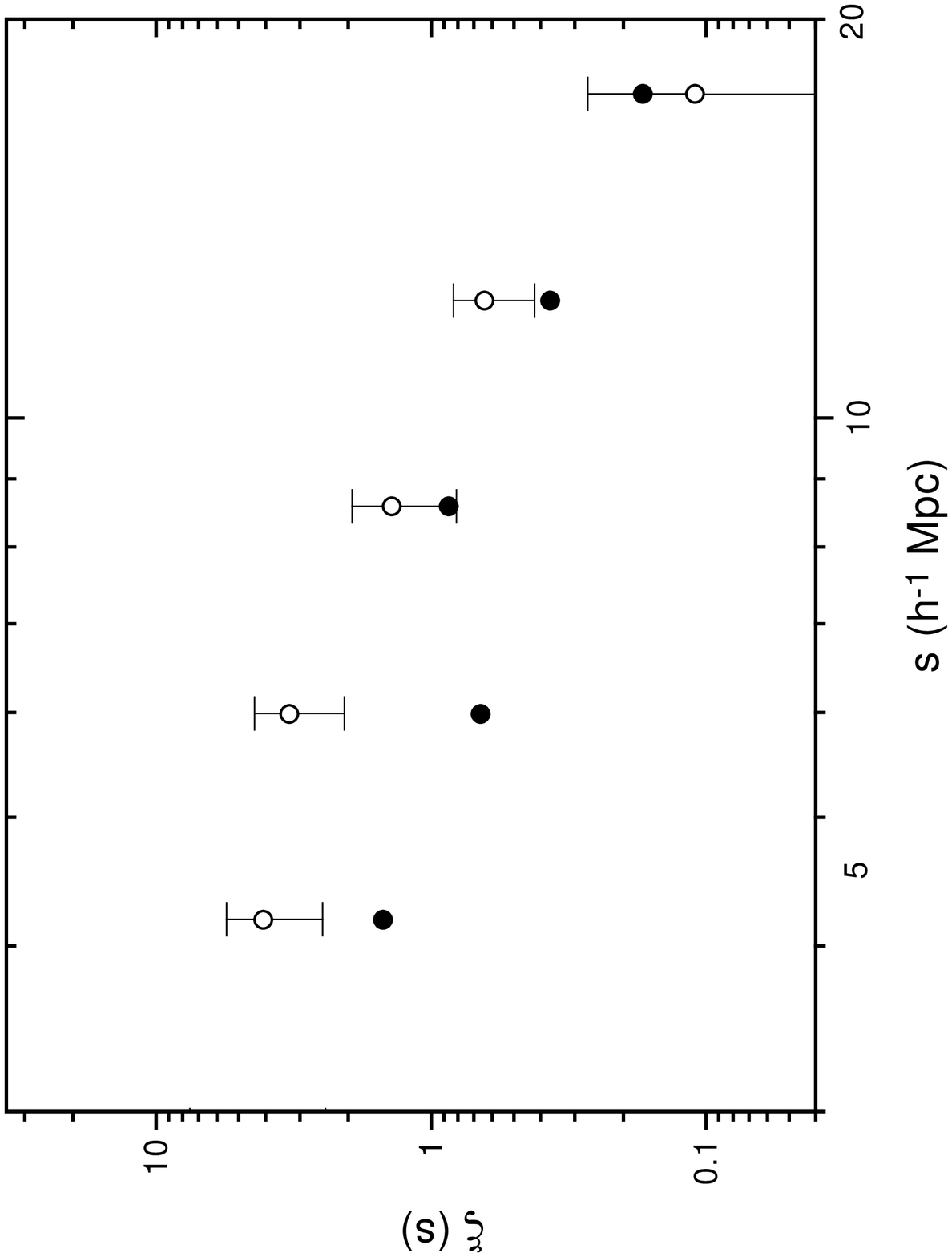]{Comparison between the correlation functions 
for all P groups with $n\geq 3$ members having velocity dispersion 
greater (open circles) and smaller (dots) than 
$\sigma_v=129$ km/s.}

\figcaption[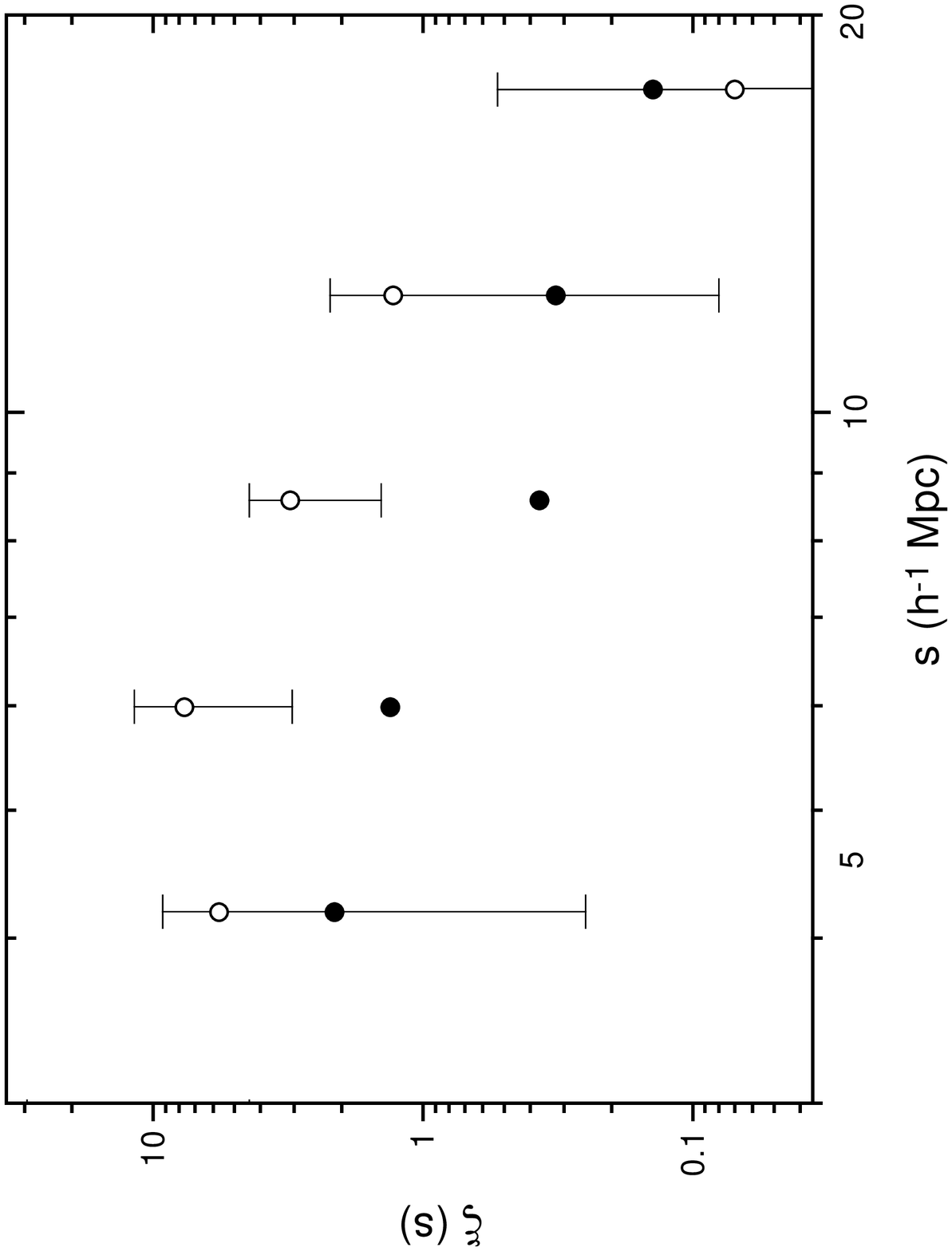]{Comparison between the correlation functions 
for the volume-limited sample (with depth of 4000 km/s) of the H 
groups with $n\geq 4$ members having projected virial radii 
greater (open circles) and smaller (dots) 
than $R_{pv}=0.9\;h^{-1}$ Mpc.}

\figcaption[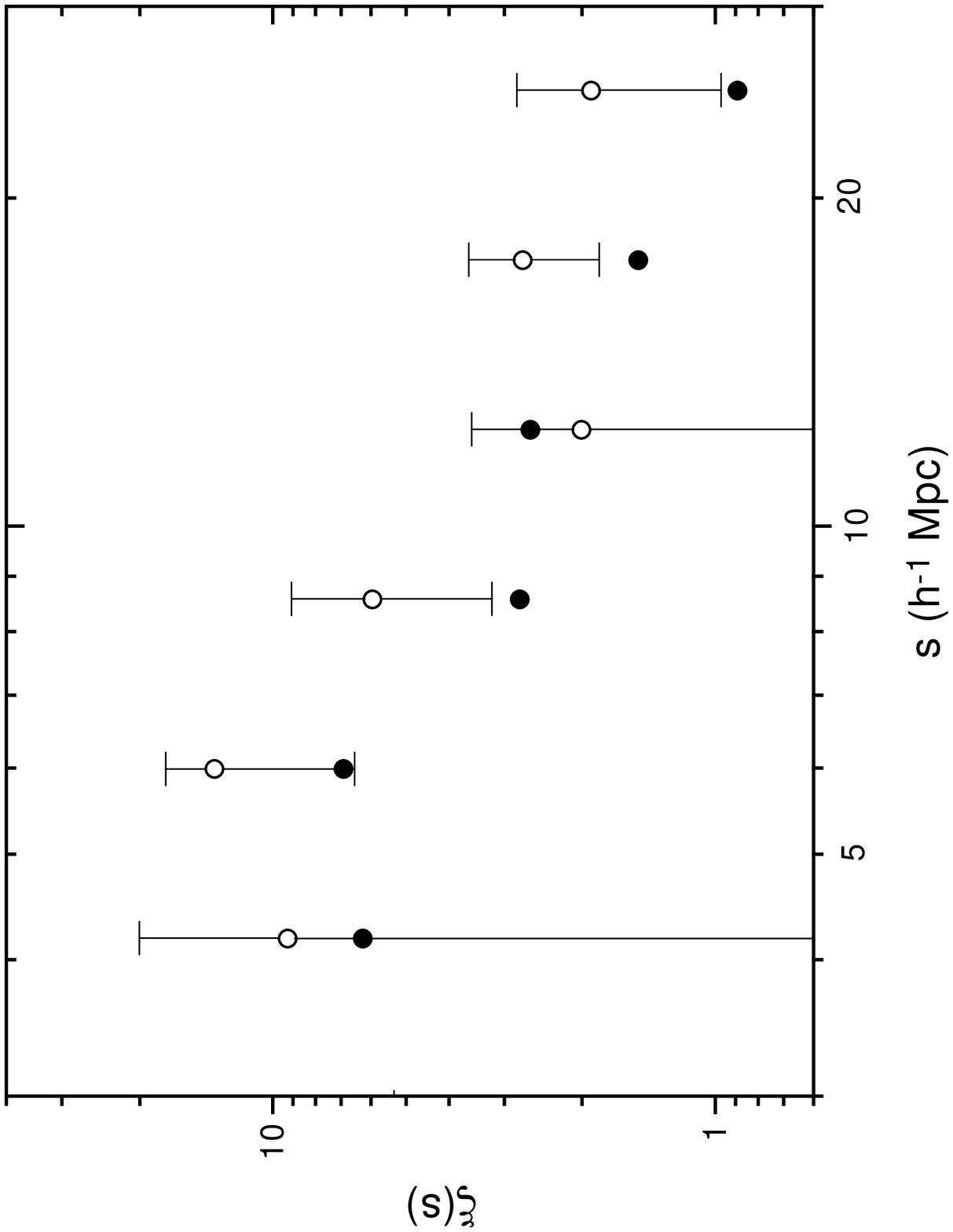]{Comparison between the correlation functions 
for the volume-limited sample (with depth of 4000 km/s) of the 
P groups with $n\geq 4$ members having virial masses greater (open 
circles) and smaller (dots) than 
$M_v=2.7\cdot 10^{13}\;h^{-1}\;M_{\odot}$.}

\figcaption[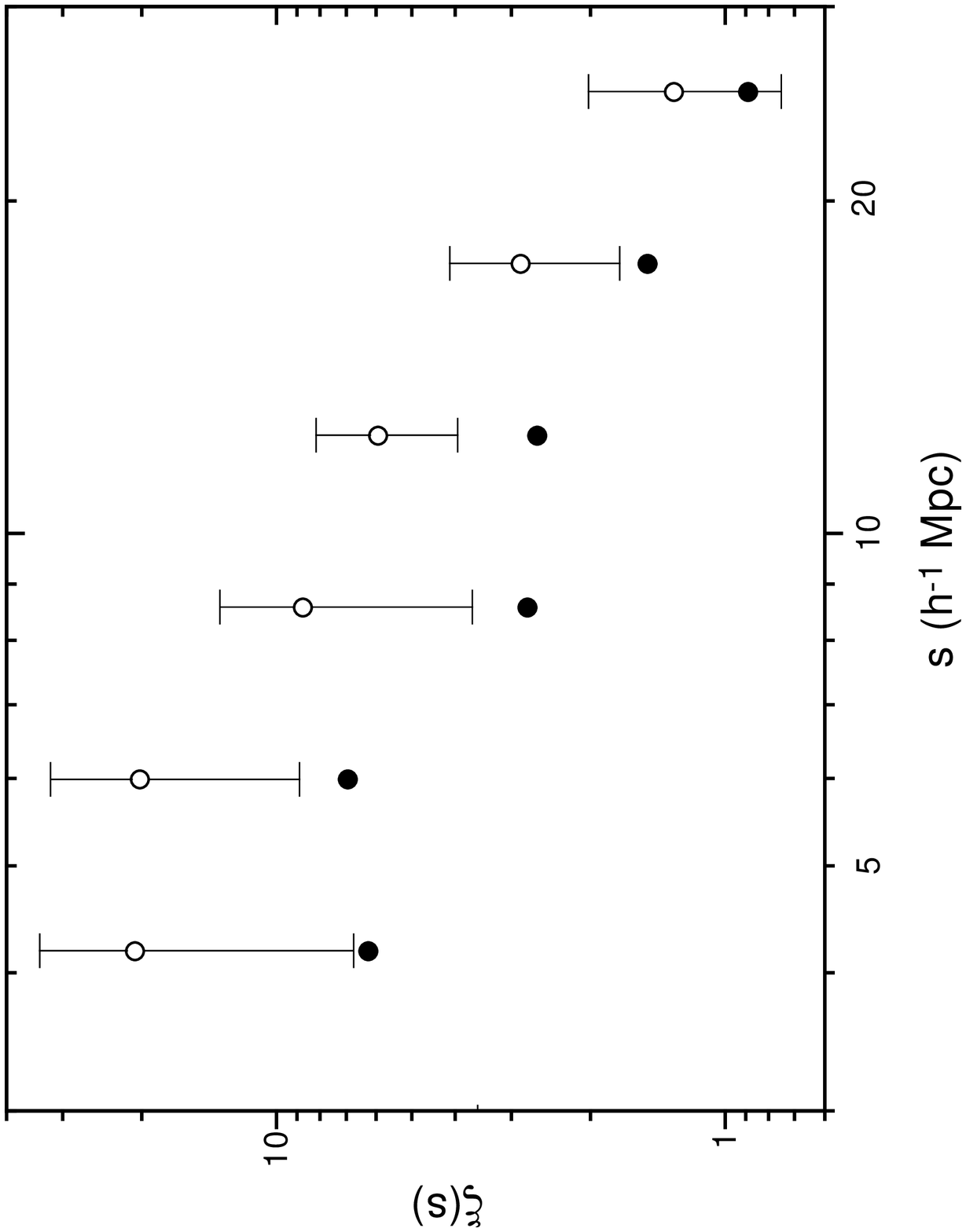]{Comparison between the correlation functions 
for the volume-limited sample (with depth of 4000 km/s) of the H 
groups with $n\geq 4$ members having virial masses greater (open 
circles) and smaller (dots) than 
$M_v=1.5\cdot 10^{13}\;h^{-1}\;M_{\odot}$.}. 

\newpage
\clearpage

\begin{deluxetable}{lrcccc}

\tablewidth{0pc}
\tablecaption{Correlation function parameters for different morphologies from the whole NOG}
\tablehead{
\colhead{type} &
\colhead{N} &
\colhead{range} &
\colhead{$s_0$} &
\colhead{$\gamma$} &
\colhead{$\sigma_8$}\\
\colhead{ } & 
\colhead{ } &
\colhead{($h^{-1}$Mpc)} &
\colhead{($h^{-1}$Mpc)} & 
\colhead{ } & 
\colhead{ }\\
}
\startdata
E-S0 & 1036 & 2.7 - 12 & $11.1\pm0.5$   & $1.5\pm0.1$   & $1.6\pm0.1$\\
Sp   & 5899 & 2.7 - 12 & $ 5.6\pm0.1  $ & $1.49\pm0.07$ & $0.94\pm0.02$\\
E    &  621 & 2.7 - 12 & $ 9.7\pm1.0$   & $1.8\pm0.3$   & $1.6\pm0.3$\\
S0   &  415 & 2.7 - 12 & $10.9\pm1.3$   & $1.6\pm0.3$   & $1.7\pm0.3$\\
S0/a &  496 & 2.7 - 17 & $8.5\pm0.9$    & $1.2\pm0.2$   & $1.2\pm0.1$\\
Sa   &  845 & 2.7 - 12 & $7.4\pm0.5$    & $1.5\pm0.2$   & $1.2\pm0.1$\\
Sb   &  822 & 2.7 - 12 & $5.5\pm0.3$    & $1.8\pm0.3$   & $1.0\pm0.1$\\
Scd  & 3736 & 2.7 - 12 & $5.1\pm0.2$    & $1.5\pm0.1$   & $0.88\pm0.03$\\

\enddata
\end{deluxetable}
\clearpage

\begin{deluxetable}{lrcccc}

\tablewidth{0pc}
\tablecaption{Correlation function parameters for different morphologies from the volume-limited sample at 6000 km/s}
\tablehead{
\colhead{type} &
\colhead{N} &
\colhead{range} &
\colhead{$s_0$} &
\colhead{$\gamma$} &
\colhead{$\sigma_8$}\\
\colhead{ } & 
\colhead{ } &
\colhead{($h^{-1}$Mpc)} &
\colhead{($h^{-1}$Mpc)} & 
\colhead{ } & 
\colhead{ }\\
}
\startdata
E-S0 &  407 & 2.7 - 12 & $11.6\pm0.8$   & $1.5\pm0.2$   & $1.6\pm0.2$\\
Sp   & 1826 & 2.7 - 12 & $ 6.4\pm0.2$   & $1.5\pm0.1$   &$ 1.04\pm0.03$\\
E    &  297 & 2.7 - 12 & $10.5\pm1.0$   & $1.6\pm0.2$   & $1.6\pm0.2$\\
S0   &  110 & 2.7 - 12 & $11.0\pm1.6$   & $2.1\pm0.4$   & $2.2\pm0.8$\\
Sa   &  308 & 2.7 - 12 & $ 9.0\pm1.0$   & $1.3\pm0.3$   & $1.3\pm0.1$\\
Sb   &  358 & 2.7 - 12 & $ 6.0\pm0.4$   & $1.9\pm0.3$   & $1.1\pm0.1$\\
Scd  & 1059 & 2.7 - 12 & $ 5.5\pm0.2$   & $1.4\pm0.1$   & $0.92\pm0.02$\\

\enddata
\end{deluxetable}
\clearpage

\begin{deluxetable}{lrcccccc}

\tablewidth{0pc}
\tablecaption{Correlation function parameters for different luminosity intervals}
\tablehead{
\colhead{sample} &
\colhead{N} &
\colhead{range} &
\colhead{$s_0$} & 
\colhead{$\gamma$} &
\colhead{$\sigma_8$} &
\colhead{$n_g$} &
\colhead{d}\\
\colhead{ } & 
\colhead{ } &
\colhead{($h^{-1}$Mpc)} & 
\colhead{($h^{-1}$Mpc)} &
\colhead{ } &
\colhead{ } &
\colhead{($h^3$Mpc$^{-3}$)} &
\colhead{($h^{-1}$Mpc)}\\
}
\startdata
$M_B\leq -21.0$  & 199 & 2.7 - 12 & $11.9\pm1.9$ & $1.3\pm0.3$ & $1.5\pm0.3$  & $3.4\cdot 10^{-4}$ & 14\\
$M_B\leq -20.6$  & 584 & 2.7 - 12 & $ 8.7\pm0.3$ & $1.6\pm0.1$ & $1.4\pm0.1$  & $1.0\cdot 10^{-3}$ & 10\\
$M_B\leq -20.3$  &1119 & 2.7 - 12 & $ 7.8\pm0.2$ & $1.5\pm0.1$ & $1.21\pm0.04$ & $1.9\cdot 10^{-3}$ &  8\\
$M_B\leq -19.89$ &2257 & 2.7 - 12 & $ 7.2\pm0.2$ & $1.5\pm0.1$ & $1.14\pm0.03$ & $3.9\cdot 10^{-3}$ &  6\\

\enddata
\end{deluxetable}
\clearpage

\begin{deluxetable}{lrc}

\tablewidth{0pc}
\tablecaption{The values of $<r>$ for different samples of groups}
\tablehead{
\colhead{sample} & 
\colhead{N} &
\colhead{$<r>$}\\
}
\startdata
H groups ($n\geq 3$) & 140 & $1.19\pm0.34$\\
P groups ($n\geq 3$) & 141 & $1.26\pm0.52$\\
H groups ($n\geq 4$) &  78 & $2.10\pm1.03$\\
P groups ($n\geq 4$) &  83 & $1.65\pm0.94$\\
H groups ($n\geq 5$) &  50 & $2.23\pm1.30$\\
P groups ($n\geq 5$) &  53 & $2.27\pm0.78$\\
H groups ($n\geq 6$) &  36 & $3.14\pm1.34$\\
P groups ($n\geq 6$) &  31 & $2.23\pm1.83$\\
H groups ($n\geq 7$) &  25 & $3.62\pm2.62$\\
P groups ($n\geq 7$) &  22 & $3.62\pm2.04$\\
\enddata
\end{deluxetable}
\clearpage

\begin{deluxetable}{llrc}

\tablewidth{0pc}
\tablecaption{Correlation function amplitudes for galaxies and groups}
\tablehead{
\colhead{sample} &
\colhead{flag} &
\colhead{N} &
\colhead{$\xi(s^*)$}\\
}
\startdata
galaxies            &ml& 7028 & 0.67$^{+0.05}_{-0.06}$\\
P groups ($n\geq3$) &ml&  506 & 1.09$^{+0.21}_{-0.25}$\\
H groups ($n\geq3$) &ml&  474 & 1.17$^{+0.28}_{-0.27}$\\
H groups ($n\geq4$) &ml&  280 & 1.44$^{+0.45}_{-0.43}$\\
H groups ($n\geq5$) &ml&  189 & 1.45$^{+0.54}_{-0.57}$\\
galaxies            &vl& 1895 & 0.46$^{+0.03}_{-0.03}$\\
P groups ($n\geq3$) &vl&  141 & 0.69$^{+0.17}_{-0.18}$\\
P groups ($n\geq4$) &vl&   83 & 0.90$^{+0.29}_{-0.28}$\\
P groups ($n\geq5$) &vl&   53 & 1.07$^{+0.43}_{-0.44}$\\
P groups ($n\geq6$) &vl&   31 & 0.69$^{+0.54}_{-0.61}$\\
H groups ($n\geq3$) &vl&  140 & 0.61$^{+0.16}_{-0.16}$\\
H groups ($n\geq4$) &vl&   78 & 0.98$^{+0.34}_{-0.34}$\\
H groups ($n\geq5$) &vl&   50 & 0.96$^{+0.45}_{-0.42}$\\
H groups ($n\geq6$) &vl&   36 & 1.53$^{+0.51}_{-0.61}$\\

\enddata
\end{deluxetable}
\clearpage

\begin{deluxetable}{lccc}

\tablewidth{0pc}
\tablecaption{Correlation function amplitudes for high- and low-$\sigma_v$ groups}
\tablehead{
\colhead{sample} &
\colhead{$\sigma_v^{\dag}$ (km/s)} &
\colhead{N} &
\colhead{$\xi(s^{*}$)}\\
}
\startdata
P groups ($n\geq3$, low  $\sigma_v$) & 136.2 & 71  & 0.46$^{+0.20}_{-0.23}$\\
P groups ($n\geq3$, high $\sigma_v$) &   "   & 70  & 0.79$^{+0.31}_{-0.30}$\\
P groups ($n\geq4$, low  $\sigma_v$) & 150.2 & 42  & 0.78$^{+0.51}_{-0.53}$\\
P groups ($n\geq4$, high $\sigma_v$) &   "   & 41  & 1.17$^{+0.47}_{-0.40}$\\
H groups ($n\geq3$, low  $\sigma_v$) & 100.5 & 70  & 0.35$^{+0.26}_{-0.24}$\\
H groups ($n\geq3$, high $\sigma_v$) &   "   & 70  & 0.99$^{+0.35}_{-0.42}$\\
H groups ($n\geq4$, low  $\sigma_v$) & 113.2 & 42  & 0.34$^{+0.31}_{-0.34}$\\
H groups ($n\geq4$, high $\sigma_v$) &   "   & 36  & 2.55$^{+0.84}_{-0.72}$\\
\enddata
\tablecomments{We denote by $\sigma_v^{\dag}$ the limiting value adopted for subdividing 
each sample of groups into two subsamples of low and high $\sigma_v$.}
\end{deluxetable}

\clearpage

\begin{deluxetable}{lccc}

\tablewidth{0pc}
\tablecaption{Correlation function amplitudes for high- and low-size groups}
\tablehead{
\colhead{sample} &
\colhead{$R^{\dag}$ ($h^{-1}$Mpc)} &
\colhead{N} & 
\colhead{$\xi(s^{*})$}\\
}
\startdata 
P groups ($n\geq3$, low  $R_{pv}$)  & 0.54  & 71 & 0.49$^{+0.25}_{-0.26}$\\
P groups ($n\geq3$, high $R_{pv}$)  &  "    & 70 & 0.64$^{+0.41}_{-0.35}$\\
P groups ($n\geq4$, low  $R_{pv}$)  & 0.56  & 42 & 0.56$^{+0.32}_{-0.36}$\\
P groups ($n\geq4$, high $R_{pv}$)  &  "    & 41 & 1.44$^{+0.56}_{-0.61}$\\
P groups ($n\geq4$, low  $R_{pv}$)  & 0.70  & 50 & 0.66$^{+0.36}_{-0.41}$\\
P groups ($n\geq4$, high $R_{pv}$)  &  "    & 33 & 1.70$^{+0.70}_{-0.75}$\\
H groups ($n\geq3$, low  $R_{pv}$)  & 0.66  & 70 & 0.35$^{+0.19}_{-0.20}$\\
H groups ($n\geq3$, high $R_{pv}$)  &  "    & 70 & 0.77$^{+0.31}_{-0.31}$\\
H groups ($n\geq4$, low  $R_{pv}$)  & 0.64  & 39 & 0.45$^{+0.27}_{-0.31}$\\
H groups ($n\geq4$, high $R_{pv}$)  &  "    & 39 & 1.95$^{+0.61}_{-0.72}$\\
H groups ($n\geq4$, low  $R_{pv}$)  & 0.90  & 58 & 0.61$^{+0.33}_{-0.31}$\\
H groups ($n\geq4$, high $R_{pv}$)  &  "    & 20 & 3.73$^{+1.27}_{-1.16}$\\
P groups ($n\geq3$, low  $R_m$)     & 0.49  & 71 & 0.49$^{+0.21}_{-0.24}$\\
P groups ($n\geq3$, high $R_m$)     &  "    & 70 & 0.76$^{+0.27}_{-0.33}$\\
P groups ($n\geq4$, low  $R_m$)     & 0.60  & 42 & 0.74$^{+0.48}_{-0.47}$\\
P groups ($n\geq4$, high $R_m$)     &  "    & 41 & 1.62$^{+0.68}_{-0.73}$\\
P groups ($n\geq4$, low  $R_m$)     & 0.76  & 55 & 0.65$^{+0.66}_{-0.63}$\\
P groups ($n\geq4$, high $R_m$)     &  "    & 28 & 1.27$^{+0.66}_{-0.63}$\\
H groups ($n\geq3$, low  $R_m$)     & 0.74  & 71 & 0.46$^{+0.26}_{-0.25}$\\
H groups ($n\geq3$, high $R_m$)     &  "    & 70 & 1.00$^{+0.35}_{-0.33}$\\
H groups ($n\geq4$, low  $R_m$)     & 0.78  & 39 & 0.89$^{+0.52}_{-0.53}$\\
H groups ($n\geq4$, high $R_m$)     &  "    & 39 & 1.46$^{+0.60}_{-0.59}$\\
H groups ($n\geq4$, low  $R_m$)     & 1.0   & 55 & 0.83$^{+0.39}_{-0.39}$\\
H groups ($n\geq4$, high $R_m$)     &  "    & 23 & 2.39$^{+1.11}_{-1.05}$\\
\enddata
\tablecomments{We denote by $R^{\dag}$ the limiting value 
adopted for subdividing each sample of groups into two subsamples of high and low 
size.}
\end{deluxetable}

\clearpage

\begin{deluxetable}{lccc}

\tablewidth{0pc}
\tablecaption{Correlation function amplitudes for high- and low-mass groups}
\tablehead{
\colhead{sample} &
\colhead{$M_v^{\dag}$ ($h^{-1}\;M_{\odot}$)} &
\colhead{N} & 
\colhead{$\xi(s^{*})$}\\
}
\startdata
P groups ($n\geq3$, low $M_v$)  & $1.17\cdot 10^{13}$ & 68 & 0.58$^{+0.28}_{-0.28}$\\
P groups ($n\geq3$, high $M_v$) &         "           & 67 & 0.65$^{+0.28}_{-0.30}$\\
P groups ($n\geq4$, low $M_v$)  & $1.41\cdot 10^{13}$ & 41 & 0.65$^{+0.47}_{-0.50}$\\
P groups ($n\geq4$, high $M_v$) &         "           & 41 & 1.17$^{+0.46}_{-0.50}$\\
P groups ($n\geq4$, low  $M_v$) & $2.70\cdot 10^{13}$ & 55 & 0.77$^{+0.38}_{-0.38}$\\
P groups ($n\geq4$, high $M_v$) &         "           & 27 & 1.22$^{+0.57}_{-0.64}$\\
H groups ($n\geq3$, low $M_v$)  & $6.86\cdot 10^{12}$ & 71 & 0.40$^{+0.28}_{-0.25}$\\
H groups ($n\geq3$, high $M_v$) &         "           & 69 & 1.26$^{+0.39}_{-0.39}$\\
H groups ($n\geq4$, low $M_v$)  & $9.20\cdot 10^{12}$ & 38 & 0.43$^{+0.30}_{-0.38}$\\
H groups ($n\geq4$, high $M_v$) &         "           & 38 & 2.60$^{+0.76}_{-0.78}$\\
H groups ($n\geq4$, low $M_v$)  & $1.5\cdot  10^{13}$ & 49 & 0.46$^{+0.27}_{-0.23}$\\
H groups ($n\geq4$, high $M_v$) &         "           & 27 & 2.68$^{+0.95}_{-0.93}$\\

\enddata
\tablecomments{We denote by $M_v^{\dag}$ the limiting value adopted for 
subdividing each sample of groups into two subsamples of high and low $M_v$.}
\end{deluxetable}

\clearpage

\begin{figure}
\centerline{
\psfig{figure=fig1new.eps,height=15cm,angle=270}
}
\end{figure}
\clearpage

\begin{figure}
\centerline{
\psfig{figure=fig2new.eps,height=15cm,angle=270}
}
\end{figure}
\clearpage

\begin{figure}
\centerline{
\psfig{figure=fig3new.eps,height=15cm,angle=270}
}
\end{figure}
\clearpage

\begin{figure}
\centerline{
\psfig{figure=fig4new.eps,height=15cm,angle=270}
}
\end{figure}
\clearpage

\begin{figure}
\centerline{
\psfig{figure=fig5new.eps,height=15cm,angle=270}
}
\end{figure}
\clearpage

\begin{figure}
\centerline{
\psfig{figure=fig6new.eps,height=15cm,angle=270}
}
\end{figure}
\clearpage

\begin{figure}
\centerline{
\psfig{figure=fig7new.eps,height=15cm,angle=270}
}
\end{figure}
\clearpage

\begin{figure}
\centerline{
\psfig{figure=fig8new.eps,height=15cm,angle=270}
}
\end{figure}
\clearpage

\begin{figure}
\centerline{
\psfig{figure=fig9new.eps,height=15cm,angle=270}
}
\end{figure}
\clearpage

\begin{figure}
\centerline{
\psfig{figure=fig10new.eps,height=15cm,angle=270}
}
\end{figure}
\clearpage

\begin{figure}
\centerline{
\psfig{figure=fig11new.eps,height=15cm,angle=270}
}
\end{figure}
\clearpage

\begin{figure}
\centerline{
\psfig{figure=fig12new.eps,height=15cm,angle=270}
}
\end{figure}
\clearpage

\begin{figure}
\centerline{
\psfig{figure=fig13new.eps,height=15cm,angle=270}
}
\end{figure}
\clearpage

\begin{figure}
\centerline{
\psfig{figure=fig14new.eps,height=15cm,angle=270}
}
\end{figure}
\clearpage

\begin{figure}
\centerline{
\psfig{figure=fig15new.eps,height=15cm,angle=270}
}
\end{figure}
\clearpage

\begin{figure}
\centerline{
\psfig{figure=fig16new.eps,height=15cm,angle=270}
}
\end{figure}

\end{document}